\documentstyle[12pt]{article}
\def\be{\begin{equation}}
\def\ee{\end{equation}}
\def\bea{\begin{eqnarray}}
\def\eea{\end{eqnarray}}
\def\beano{\begin{eqnarray*}}
\def\eeano{\end{eqnarray*}}

\long\def \omitThis #1 {}
\def\sc{\scriptstyle}
\def\scsc{\scriptscriptstyle}
\def\sfrac#1#2{ {\sc \frac{{#1}}{{#2}} } }
\def\ts  {\thinspace}
\def\seq {\! = \!}
\def\q   {{\bf q}}
\def\r   {{\bf r}}
\def\qh  {{\hat{q}}}
\def\qs  {{q^\star}}
\def\qsd {q_{\Delta}^\star}
\def\alp {\alpha}
\def\ah  {\hat{\alp}}
\def\aq  {\alp_{{\sc V}}(q)}
\def\aqs {\alp_V(\qs)}
\def\aF  {\alp_F(r)}
\def\eps {\epsilon}
\def\sig {\sigma}
\def\qos {q_0/\protect\sqrt{\sig}}
\def\rs  {r_0 \protect\sqrt{\sig}}
\def\Lam {\Lambda}
\def\lam {\lambda}
\def\Var {V_{a^2}({\bf r})}
\def\Va  {V_{a^2}}
\def\eIR {e_{{\rm \scsc IR}}}
\def\e   {{\rm e}}
\def\plaq{{\rm plaq}}

\def\O   {{\cal O}}
\def\MSbar { \overline{{\rm MS}} }

\def\no{\noindent}
\def\ie{{\it i.e.}}
\def\eg{{\it e.g.}}

\begin{document}

\pagestyle{empty}

\begin{flushright}
CLNS 94/1294 (revised) \\
hep-lat/9408016 \\
August 1994 \\
\end{flushright}

\bigskip
\bigskip

\begin{center}
{\Large \bf The (Lattice) QCD Potential and Coupling: \\
\vskip 1.9mm
How to Accurately Interpolate Between \\
\vskip 2.3mm
Multi-Loop QCD and the String Picture \\
}

\vskip 17mm

Timothy R. Klassen \\
\vskip 3mm
\small Newman Laboratory of Nuclear Studies \\
\vskip 1mm
\small Cornell University \\
\vskip 1mm
\small Ithaca, NY 14853 \\
\end{center}

\vskip 9mm

\centerline{{\bf Abstract}}
\bigskip

\no
We present a simple parameterization of the running coupling constant
$\aq$, defined via the static potential, that interpolates between
2-loop QCD in the UV and the string prediction
{}~$V(r)=\sig r - \frac{\pi}{12 r}$~ in the IR.
Besides the usual $\Lam$-parameter and the string tension $\sig$,
$\aq$ depends on one dimensionless parameter, determining how fast the
crossover {}from UV to IR behavior occurs (in principle we know how to
take into account any number of loops by adding more parameters).
Using a new ansatz for the {\it lattice} potential in terms of the
{\it continuum} $\aq$, we can fit quenched and unquenched
Monte Carlo results for the potential down to {\it one} lattice spacing,
%% without introducing any extra parameters to accommodate lattice
%% artifacts.
and at the same time extract $\aq$ to high precision.
We compare our ansatz with
%% after PRDrev: the Heller-Karsch
1-loop results for the
lattice potential, and use $\aq$ {}from our fits to
quantitatively check the accuracy of 2-loop evolution, compare with
the Lepage-Mackenzie estimate of the coupling extracted {}from the
plaquette, and determine Sommer's scale $r_0$ much more accurately than
previously possible.  For pure SU(3) we find that $\aq$ scales on the
percent level for $\beta\geq 6$.

\bigskip\bigskip

\no

\clearpage

\pagestyle{plain}
\pagenumbering{arabic}

\section{Introduction}\label{intro}

In a situation where quarks are non-relativistic their dynamics is much
slower than that of the massless gluons. In the limit of infinitely
heavy quarks, or, equivalently, static sources, their interaction can be
described by a local, instantaneous potential\footnote{The situation is
more subtle in the presence of a non-perturbative gluon condensate.
But it is now clear that the presence of a gluon condensate is not
necessarily incompatible with the existence of a local potential.
See~\cite{GromRev} for a review and references.}
$V(r)$.
This {\it static potential} will for the
purposes of this paper always refer to the
potential between a quark and an anti-quark in the fundamental
representation of SU($N$).

For very short distances, $V(r)$ is Coulomb like, with a running
charge that is known to two loops in terms of the (unknown)
$\Lam$-parameter.  The strong coupling expansion of lattice gauge
theory suggests a string picture, where a flux tube connects two widely
separated quarks, in accord with the idea of confinement~\cite{Wilson}
(we here ignore the string breaking in the presence of dynamical
fermions).  In addition to a linear term involving the string
tension $\sig$,
the potential has a universal large $r$
correction~\cite{LSW} due to the zero point energy of the
transverse fluctuations of the flux tube, so that in the IR
\be\label{VIR} V(r) ~=~
\sig r ~-~ \frac{\eIR}{r} ~+~ \ldots,
\ee
up to a non-universal constant.  The ``IR charge'' $\eIR=\pi (D-2)/24$
in $D$ space-time dimensions, assuming the transverse fluctuations are
described by the simplest bosonic string
theory.\footnote{Nowadays~\cite{BCN} this universal correction is
easily recognized as the Casimir energy of the (bosonic) conformal
field theory in a finite geometry with free boundary conditions that
describes the transverse fluctuations of the flux tube; $D-2$ being
the central charge of this theory.}

At present the only way to obtain quantitative information on $V(r)$
is through the Monte Carlo simulation of Wilson loops.  These are
nowadays very precise; the lattice potential can be measured with a
relative accuracy approaching $10^{-4}$ at short distances and
$10^{-3}$ at long distances. The lattices used are typically of size
$32^4$ for pure gauge theory and $16^4$ in the presence of dynamical
quarks. The spacing of the finest lattices available for SU(3) is
about 0.04~fm, corresponding to 5~GeV, just where 2-loop perturbation
theory is expected to become good. However, at these distances the
potential suffers {}from large lattice artifacts, which obscure the view
of perturbative QCD. These lattice artifacts are on the order of
$10$\% at one lattice spacing if one uses the standard (unimproved)
Wilson action for the gluons, and fall roughly like $1/r^2$ at
larger distances. One therefore has to go to many lattice spacings
before rotational invariance is restored within errors.

In early studies, the lattice potential, at least restricted to
on-axis points, was fitted to an ansatz of
the form $-e/r + \sig r + V_0$, \ie~it was assumed to behave
as in the IR, ~eq.~(\ref{VIR}). Not just does this ansatz completely
ignore the lattice artifacts, it strictly speaking contradicts QCD
at short distances. The number $e$ has to be
interpreted as some kind of effective average coupling. It is not even
clear that $e$ is an effective short-distance quantity, since it
usually comes out rather close to the IR charge $\eIR = \pi/12$ in
eq.~(\ref{VIR}).

To partially accommodate lattice effects, Michael~\cite{fitMich} has
suggested to add a term to this ansatz, proportional to the difference
between the lattice and the continuum Coulomb potential.  This takes
into account at least the tree-level lattice artifacts.
Also, since recent Monte Carlo data~\cite{fitMich,UKsutwo,BS} on fine
lattices are accurate
enough to see that $e$ really is a running charge, Michael~\cite{fitMich}
proposed  to take this into account in a phenomenological way
by replacing $e$ with $e-f/r$, with some new parameter $f$.
If necessary,
we will distinguish these kind of ans\"atze from the naive
Coulomb $+$ linear one, by calling them  ``modified Coulomb $+$ linear
ans\"atze'' (or fits).

With this setup one can include smaller $r$-values in the fit, but the
first few still have to be left out --- and more and more as the data
get more precise and the lattice spacings smaller. Furthermore, the
short-distance parameters are not very stable with respect to varying
the fit range, in particular on fine lattices.  This is not
surprising, since asymptotic freedom is not properly incorporated in
this ansatz; the short-distance parameters are all effective
parameters.  Therefore, the only physical result one obtains {}from
these fits is, at best, an (improved) estimate of the string tension
$\sig$; no information about short-distance QCD, \eg~the
$\Lam$-parameter, is gained.

To learn about $\Lam$ another idea of Michael~\cite{fitMich} is now
widely used. Namely, one defines a running coupling $\alp_F(r)$
in the  ``force scheme'' by
\be\label{alpforce}
 r^2 V'(r) ~\equiv~ C_F \alp_F(r) ~.
\ee
Here $C_F = (N^2 -1)/2N$ is the Casimir invariant of the fundamental
representation of SU($N$). One then fits numerical derivatives of the
potential data to a 2-loop formula for $\alp_F(r)$. Lattice artifacts,
errors {}from taking numerical derivatives,  and the fact that
2-loop evolution is only
% beginning to become good,
on the verge of becoming reliable,
lead to quite
large errors in $\Lam$. Furthermore, it is hard to estimate the
systematic errors of this $\Lam$-determination; one might simply
be determining an {\it effective} $\Lam$, that is still quite
far {}from the true $\Lam$.
This could be true even if some consistency checks involving
different lattice spacings, \eg~scaling of quantities involving $\Lam$,
work quite well. In fact, our results indicate that this is exactly
what happens.

We here describe a new method that allows for a unified fit of
potential data at all distances by incorporating lattice artifacts
and the running of the coupling in a more fundamental way. The
general idea is to express to the lattice potential in terms of a
running coupling. More precisely there are three ingredients:

\begin{enumerate}

\item
As advocated in refs.~\cite{BuGrTy,BLM}, for example, use the continuum
static potential to define a physical running coupling $\aq$ via
\be\label{alpofV} V(q) ~
\equiv ~ -C_F ~\frac{4\pi \aq}{q^2} ~.
\ee
This scheme for the coupling will be referred to as the V scheme.
With this definition the effect of multi-gluon exchange is absorbed
into $\aq$.  Fourier transforming eq.~(\ref{alpofV}), the continuum
potential is by definition given by
\be\label{Vofalp} V(r) ~=~ - C_F ~\int_{-\infty}^{\infty}
{}~\frac{d^3q}{(2\pi)^3}~ {\rm e}^{-i \q \r}~\frac{4\pi \aq}{q^2} ~=~ -
\frac{2 C_F}{\pi}~ \int_0^{\infty} ~dq ~\frac{\sin q r}{q r} ~\aq ~.
\ee

\item
Globally parameterize $\aq$ to take into account 2-loop QCD in the UV
in terms of $\Lam$, the string prediction of eq.~(\ref{VIR}) in the IR
in terms of $\sig$, and to have another parameter
determining how fast the crossover {}from the UV to the IR behavior
occurs.  [Actually we will see that by introducing further parameters
we can take into account any number of loops in the UV.]

\item
Finally, make the ansatz that in terms
of the {\it continuum} $\aq$ the potential on an infinite
{\it lattice} of spacing
$a$ can be accurately represented as,
\be\label{Vaofalp} \Var ~=~  V_0 ~-~ C_F
{}~\int_{-\pi/a}^{\pi/a} ~\frac{d^3q}{(2\pi)^3} ~{\rm e}^{-i \q
\r}~\frac{4\pi \alp_V(\hat{q})}{\hat{q}^2} ~, ~~~ ~~~~ \hat{q}_i
\equiv \frac{2}{a}\sin\frac{a q_i}{2} ~,
\ee
except presumably for $r=0$, where the whole underlying picture of
gluon exchange breaks down (which is why we have to add the constant
$V_0$ after making the integral convergent by subtracting its value
at $r=0$).\footnote{The ``lattice momentum'' $\hat{q}$ in
eq.~(\ref{Vaofalp}) is appropriate for the usual Wilson action.
If an improved action is used, $\hat{q}$ should be replaced by
the corresponding ``improved momentum'', yielding an improved
potential $V_{a^4}({\bf r})$, {\it etc.}}

\end{enumerate}

\no
The three parameters in $\aq$ and the constant $V_0$ are then fitted
by matching eq.~(\ref{Vaofalp}) to the Monte Carlo data for the
potential.

To be sure, the above equation for the lattice potential in terms of
the continuum coupling must be regarded as an ansatz.
The motivation for this ansatz is that it appears to be the most
natural expression that (a) gives the exact tree level lattice potential,
that is, the lattice Coulomb potential (corresponding to the case
$\aq \equiv {\rm const}$),
and (b) has the  right continuum limit.
The fact that eq.~(\ref{Vaofalp}) will allow us to obtain excellent
fits of  the lattice potential of a wide variety of theories
down to {\it one} lattice spacing,
provides indirect empirical evidence for its accuracy. Additional
evidence will be discussed later.

The outline of this paper is as follows. In sect.~2 we provide the
details of the parameterization of $\aq$ satisfying the required UV
and IR behavior.  In particular, we prove an elementary theorem showing
how the running coupling can be written in terms of ``iterated
logarithms'' to all orders of the loop expansion of the
$\beta$-function. This theorem then suggests a natural way of avoiding
the perturbative Landau pole, motivating our ansatz for the running
coupling.

In sect.~3 we discuss the efficient evaluation of the
three-dimensional integral in eq.~(\ref{Vaofalp}), and details of our
fitting and error analysis procedure.
In sect.~4 we apply our results to Monte Carlo data for the lattice
potential of various gauge theories; with gauge group SU(3) or
SU(2), with or without dynamical fermions (Wilson and staggered).
Using eq.~(\ref{Vaofalp}) leads to excellent fits of the data down
to one lattice spacing.
To check for systematic errors in the running coupling $\aq$ extracted
{}from these fits, we perform fits with an $\aq$ that partially
incorporates 3-loop effects. We compare our results with the
Lepage-Mackenzie estimate~\cite{LM} of $\aq$ in the UV, in particular
also on very fine lattices.  We then compare the 1-loop expansion of
eq.~(\ref{Vaofalp}) with the 1-loop results of~\cite{HelKar}
%% After PRDrev:
%% Heller and Karsch~\cite{HelKar}
for the on-axis lattice potential. Curiously
enough, our 1-loop expansion agrees much better with the Monte Carlo
data than the results of~\cite{HelKar}. These comparisons provide good
evidence for the accuracy of representing the lattice potential as in
eq.~(\ref{Vaofalp}).
Finally, we check how accurate 2-loop evolution is compared to our
non-perturbative $\aq$.

Sect.~5 is devoted to scaling. We use the intermediate scale
$r_0$ introduced  by Sommer~\cite{Som}
to determine the ratios of lattice spacings of
various theories much more accurately than previously possible.
For pure SU(3) we find that $\aq$ scales on the percent level for
$\beta \geq 6.0$.
In sect.~6 we compare our approach with other schemes to define and
extract a running coupling {}from Monte Carlo simulations. In particular, we
provide some details about the force scheme and its close relation
to the V scheme. We conclude in sect.~7 by summarizing our findings
and outlining various applications and extensions.

In~\cite{myLATBi} we presented a brief summary of our results. Due to
the fact that our error analysis is somewhat involved, some of the
results given in~\cite{myLATBi} were slightly preliminary (also, in one
case we only had preliminary potential data).

\section{Parameterizing the Running Coupling}

\subsection{The Ultraviolet Regime}

We will first describe some general results concerning the integration
of the $\beta$-function
 \be\label{betafct}
 \beta(\alp) ~\equiv~ q^2 \partial_{q^2} \alpha(q) ~ = ~
 -\beta_0 \alp^2 - \beta_1 \alp^3 - \beta_2 \alp^4 ~-~ \ldots ~,
 \ee
of the coupling $\alp = g^2/4\pi$ in a generic scheme. Recall that
the 1- and 2-loop coefficients are scheme independent and equal
to~\cite{twoloop}
 \be\label{betacoeff}
 4\pi~ \beta_0 ~=~ \frac{11}{3} \ts N - \frac{2}{3}\ts n_f ~,
  ~~~~~~~
 (4\pi)^2 ~ \beta_1 ~=~ \frac{34}{3} \ts N^2 -
           \Bigg(\frac{10}{3} N + \frac{N^2 -1}{N} \Bigg)\ts n_f
 \ee
in QCD with gauge group SU($N$) and $n_f$ flavors of massless
fermions.  The higher coefficients depend on the scheme (none are
presently known in the V scheme). In terms of
$\hat{\alpha}(q) \equiv \beta_0 \alpha(q)$  the integration of the
$\beta$-function immediately leads to
\be\label{alpofalpexpn}
 \frac{1}{\ah} = t + b \ln\Bigg(\frac{1+b \ah}{\ah}\Bigg)
  - \int_0^{\ah} dx
   \Bigg(\frac{1}{\hat{\beta}(x)}+\frac{1}{x^2 (1+ b x)} \Bigg) =
  t + b \ln\frac{1}{\ah} + b_2 \ah + b_3 \ah^2 + \ldots,
\ee
where $\hat{\beta}(\ah) \equiv
      -\ah^2 (1+\hat{\beta}_1 \ah + \hat{\beta}_2 \ah^2 + \ldots)$,
{}~$\hat{\beta}_i \equiv \beta_i/\beta_0^{i+1}$, and
\be\label{tbbb}
  t ~\equiv~ \ln(\frac{q^2}{\Lam^2}) ~,
 ~~~ b ~\equiv~ \hat{\beta}_1 ~,
 ~~~ b_2 ~\equiv~ \hat{\beta}_1^2 - \hat{\beta}_2 ~,
 ~~~ b_3 ~\equiv~ \hat{\beta}_1 \hat{\beta}_2 - \sfrac{1}{2}\hat{\beta}_1^3
                                              - \sfrac{1}{2}\hat{\beta}_3 ~.
 \ee
We have followed the usual convention of defining the integration
constant,  the $\Lam$-parameter,
so that there is no constant term on the rhs of~(\ref{alpofalpexpn}).
Note that as long as the $\beta$-function has a power series expansion
in $\alp$, there is only one logarithmic term on the rhs
of~(\ref{alpofalpexpn}), coming {}from the first two loops.
Solving~(\ref{alpofalpexpn}) iteratively gives
\be\label{alpoftexpn}
 \frac{1}{\ah} ~=~ t + b \ln t + \frac{1}{t}\Big[b^2 \ln t + b_2 \Big] +
 \frac{1}{t^2}\Big[-\frac{1}{2} b^3 \ln^2 t +(b^3-b b_2)\ln t
                    + b b_2 + b_3 \Big] + \ldots .
\ee
As an aside we would like to point out that the full 2-loop expression
for $1/\ah$ involves
not just the leading terms ~$t + b \ln t$~ usually quoted, but also
the subleading 2-loop term ~$b^2 \ln t /t$.

Recall that the above expansion has a Landau pole, \ie~the perturbative
$\ah(q)$ diverges for some $q > \Lam$ as $q$ approaches $\Lam$ {}from
above. One of the goals of the ansatz we will later use for $\aq$ is of
course to avoid this unphysical Landau pole. To motivate our ansatz we
now describe a different way of solving~(\ref{alpofalpexpn}) for
$\alpha(q)$. Namely, consider the following recursively defined function
\be\label{Rkdef}
 R(t) \equiv R^{(0)}(t)~, ~~~
 R^{(k)}(t) \equiv b \ln\Big[a_k(t+R^{(k+1)}(t))\Big] ~, ~~k=0,1,2,\ldots .
\ee
We now prove that by suitably choosing the coefficients $a_k > 0$
% the solution of~(\ref{alpofalpexpn}) is given by
the running coupling can be written as
\be\label{alpasiterlog}
 \frac{1}{\ah} ~=~  t + R(t) ~.
\ee

\no
This is equivalent to showing that $R(t) = t - \frac{1}{\ah}$ can be
rewritten in terms of $\ah$ in the form
$R=b \ln\frac{1}{\ah} + b_2 \ah + b_3 \ah^2 + \ldots$. The proof
is simple: By substituting $t=\frac{1}{\ah}-R^{(0)}$ in the definition of
$R^{(k)}$ we have
\be\label{Rkii}
  R^{(k)} ~=~ b ~\ln\Big(\frac{a_k}{\ah}\Big) +
   b ~\ln\Big[ 1~+~ \ah (R^{(k+1)} - R^{(0)})\Big] ~.
\ee
All we have to do is to iterate these equations, starting with
$R^{(k)} = b \ln(\frac{a_k}{\ah})$. By induction one therefore
proves %% RADIUS OF CONVERGENCE?
\be\label{Rkexpn}
  R^{(k)} ~=~ b ~\ln\Big(\frac{a_k}{\ah}\Big) ~+~
              b ~\sum_{n=1}^{\infty} R_n^{(k)} ~(b \ah)^n ~,
\ee
where
%% \be\label{Rkn}
%%  R_n^{(k)} = \ln a_{k+n} + {\rm terms~involving~only}~\ln a_j~{\rm with}
%%                                                 ~j < n+k ~.
%% \ee
$R_n^{(k)} = \ln a_{k+n} ~+$ terms involving only $\ln a_j$ with $j < n+k$.~
This completes the proof.

\vskip 2mm

Specifically, $R_1^{(k)} = \ln(a_{k+1}/a_0)$,
              $R_2^{(k)} = \ln(a_{k+2}/a_1) - \frac{1}{2}\ln^2(a_{k+1}/a_0)$,
etc.
So the first few $a_k$ are determined by
\be\label{ak}
 a_0 ~=~ 1 ~, ~~~ b^2 \ln a_1 ~=~ b_2~, ~~~
         b^3(\ln a_2 - \ln a_1 - \frac{1}{2} \ln^2 a_1) ~=~ b_3 ~.
\ee
Note that to obtain the $1/\ah$ expansion to $n$ loops one can truncate
the iterative definition of $R(t)$ by setting $R^{(n-1)} = b \ln t$.
For example, to get two loops we can simply set $R(t)=b \ln(t+b \ln t)$,
cf.~(\ref{alpoftexpn}).

Our iterated form of $R(t)$ still has a Landau pole, but now a simple
way to avoid it suggests itself by the replacements:
\be\label{killLPi}
 \frac{1}{\ah(q)} ~=~ t + R^{(0)}(t) ~\rightarrow~
 \frac{1}{\ah(q)} ~=~ \ln\Big[ 1 ~+~ {\rm e}^t ~{\rm e}^{R^{(0)}(t)} \Big] ~,
\ee
and
\be\label{killLPii}
 {\rm e}^{R^{(k)}(t)} ~\rightarrow~ \ln^{b}\Big[ c_k ~+~
                   {\rm e}^{a_k t} ~{\rm e}^{a_k R^{(k+1)}(t)} \Big] ~,
\ee
with some constants $c_k \geq 1$. [The constant 1 in eq.~(\ref{killLPi})
is chosen to get the right IR behavior for $\aq$, see below. In a scheme
where the coupling ``freezes out'' in the IR some constant $>1$ should
be chosen.]  Note that the
$c_k$ are non-perturbative parameters. Obviously the new $\ah(q)$ has
no Landau pole.

To two loops we can now write as ansatz for $\aq$
 \be\label{myalp}
 \frac{1}{\beta_0 \ts \aq} ~=~
  \ln\left[1 ~+~{\rm e}^t ~\ln^b[c_0 ~+~ {\rm e}^t \lam(t)] \right] ~,
              ~~~ b = \beta_1/\beta_0^2 ~, ~~ t = \ln(q^2/\Lam_V^2) ~,
 \ee
where we know that $\ln \lam(t) = R^{(1)}(t) = b \ln t$ in the UV limit.
A natural Landau pole free form of $\lam(t)$ is
 \be\label{lamoft}
 \lam(t)~=~\ln^b\left[c_1 ~+~ c ~{\rm e}^t\right] ~,
 \ee
with a new parameter $c>0$.  We will see below that $c_0$ and $c_1$ are
fixed by the required IR behavior in terms of $\Lam_V, \sig$ and
$\eIR$.  $c$ is the crossover parameter mentioned in the introduction.
% With the ansatz defined by eqs.~(\ref{myalp})
% and~(\ref{lamoft}) we therefore have three parameters in $\aq$ to fit:
% $\Lam_V, \sig$ and the cross-over parameter $c$.

\vskip 1mm

%Note that in the above we now have $t \equiv \ln q^2/\Lam_V^2$ in terms of
%the $\Lam$-parameter of the V scheme.
%
The relation of the $\Lam$-parameter in the V scheme, $\Lam_V$, to that
in another scheme is fixed by a
1-loop calculation~\cite{Cel}. For instance, the relation to the
$\MSbar$ scheme is given by~\cite{HasHas,Weisz,Fischler},
{}~$\Lam_V/\Lam_{\overline{{\rm MS}}} = \exp[(31 N-10 n_f)/(66 N-12 n_f)]$.

\vskip 1mm

If one wants to incorporate three loops, one would write
\be\label{lamoftiii}
\lam(t)  ~=~ \ln^b\left[c_1 ~+~ \e^{a_1 t} ~{\rm e}^{a_1 R^{(2)}(t)}
           \right] ~,
\ee
and for $\e^{R^{(2)}(t)}$ use the form $\ln^b(c_2 + c~{\rm e}^t)$,
say.  Note that the leading 3-loop effects are given by the
$\e^{a_1 t}$ term.  They are included if we simply replace
$\e^{a_1 R^{(2)}(t)}$ by a constant $c$.

Eq.~(\ref{myalp}) is a generalization of Richardson's
ansatz~\cite{Rich}, who writes
$\beta_0 \ts \alp_R(q) = 1/\ln[1+q^2/\Lam^2_R]$.
His potential, defined as in~(\ref{Vofalp}), has the advantage of
simplicity, containing only the single parameter $\Lam_R$. However, by
the same token, the corresponding potential can be fitted to describe
only the UV {\it or} the IR {\it or} the crossover region, but not all
three or even two of them. For potential models mainly the crossover
region is relevant, explaining why the Richardson potential works so
well in that context.

\subsection{The Infrared Regime}
To built the IR behavior given by eq.~(\ref{VIR}) into our ansatz we
have to transform eq.~(\ref{VIR}) to momentum space. {\it A priori}
the Fourier transform of ~$\sigma r$~ is ill-defined. However, the
potential corresponding to $\alp_V(q) = q^{-n}$ can be defined by
analytic continuation in $n$ to be
 \be\label{varFT}
 V(r) ~=~ -C_F ~\frac{r^{n-1}}{\Gamma(n+1) \cos\frac{n \pi}{2}}  ~,
 \ee
as long as $n$ is not an odd integer.
So~(\ref{VIR}) corresponds to the small $q$ expansion
 \be\label{alpIR}
 \aq ~=~ \frac{2}{C_F} ~\frac{\sig}{q^2} ~+~
         \frac{\eIR}{C_F} ~+~ \ldots ~,
 \ee
and matching this to  eq.~(\ref{myalp}) leads after a
small amount of algebra to
 \be\label{cIR}
 \ln^b c_0 ~=~ \frac{C_F}{2} \frac{\Lam_V^2}{\beta_0 \sig} ~, ~~~~~
 \ln^b c_1 ~=~
  \Big(1-\frac{2}{C_F}~\beta_0~ \eIR \Big) ~
                                 \frac{c_0}{2b} ~ \ln^{b+1} c_0 ~.
 \ee
For given $N$, $n_f$ and $\Lam_V$, the string tension determines $c_0$.
$c_1$ is then fixed by the IR charge. In other words, $c_0$ and $c_1$
parameterize the leading and subleading IR behavior, respectively.
This was of course already obvious {}from their definition in
eqs.~(\ref{myalp}) and~(\ref{lamoft}).

% b = 0.843 for Nc=3, Nf=0; b = 0.820 for Nf=2; b = 0.739  for Nf=4
% b = 0.843 for Nc=2, Nf=0; b = 0.806 for Nf=2; b = 0.582  for Nf=4

\subsection{Summary}

The ansatz defined by eqs.~(\ref{myalp}), (\ref{lamoft})
and~(\ref{cIR}) will be referred to as our ``standard ansatz''.  It is
consistent with 2-loop QCD in the UV and in the IR reproduces the
leading and subleading prediction of the string picture.  For given
$N$, $n_f$ and $\eIR$ we have to fit three parameters in $\aq$:
$\Lam_V$, $\sig$ and the crossover parameter $c$. The leading 3-loop
effects can be taken into
account by adding one more parameter, $a_1$ in eq.~(\ref{lamoftiii}).
At this point we fix the IR charge at
$\eIR = \frac{\pi}{12}$, leaving a check of this value to future high
precision studies.

\section{Fitting and Error Analysis}

\subsection{Numerical Evaluation of $\Var$}

It is crucial to have an efficient way of numerically evaluating $\Var$
if one wants to perform fits.
Recall first of all, that our approach is not expected to work for $r=0$.
So we can not simply subtract off the $r=0$ value of our ansatz
when comparing with the Wilson loop potential, which is normalized to
vanish for $r=0$, by such an (implicit) subtraction.\footnote{This
normalization is related to the overall constant in the potential
(defined, for instance, by the behavior of the potential in the IR).
Although this constant is physical --- it determines the
absolute energy of heavy quark bound states, for example ---
it has not yet been possible to calculate it in the Wilson loop approach,
basically because it is difficult to separate {}from self-energy
contributions. Therefore one {\it defines} the Wilson
loop potential to automatically vanish at $r\!=\!0$.
See~\cite{GromRev} for a discussion of issues related to the constant
in the potential.}
For simplicity we have therefore added a constant in
eq.~(\ref{Vaofalp}), after performing the subtraction. [The
subtraction is still useful, since then the integral has an
{\it integrable} singularity at $q=0$, and does not have to be defined
by analytic continuation.]

Since the integrand of $\Var$ is a periodic function in each $q_i$ one
can quite efficiently evaluate the integral as follows
(cf.~\cite{LWcomp,Mor}): First introduce a small gluon mass,
\ie~replace $1/\hat{q}^2$ by $1/(\hat{q}^2 + m^2)$. Next approximate
the integral by a sum; schematically
\be\label{inttosum}
 \int_{-\pi/a}^{\pi/a} d^dq ~F(q) ~~\approx~~ \Big(\frac{2\pi}{a L} \Big)^d ~
          \sum_{{\bf n}} ~F\Big(\frac{2\pi {\bf n}}{a L}\Big) ~,
\ee
(where each component of ${\bf n}$ runs {}from $-L/2$ to $L/2 -1$ in
integer steps) but only after applying the following change of
variable to the integral: $q_i \rightarrow q_i - \eps \sin q_i$, where
for given $m$, $\eps$ is determined by the equations $\eps=1/\cosh u$,
{}~$u - \tanh u = m$. Using the Poisson resummation formula and a
contour deformation argument one sees that
this choice of $\eps$ greatly
accelerates the convergence of the sum towards the integral as $L$ is
increased.
Finally, extrapolate to zero gluon mass.
For the accuracy desired here, about $10^{-5}$ for small $r$ and
$10^{-4}$ for large $r$ to be safely below the Monte Carlo errors, we
can dispense with the last step and simply use a fixed small mass such
as $a m \seq  10^{-4}$.

For small $r$ we find that $L\seq 20$ is sufficient. As the components of
${\bf r}$ increase the integral becomes more and more oscillatory, so
$L$ must increase correspondingly. If one needs to evaluate $\Var$ for
many ${\bf r}$-values, as we do, one should make sure to evaluate the
${\bf r}$-independent part of the integrand only once for a given
$2\pi {\bf n}/a L$, instead of doing so again and again for different
${\bf r}$.

For sufficiently large $r$ (depending on the accuracy of the potential
data; typically when at least one component of $\r$ is greater than
$16 a$) one can simply use the continuum $V(r)$ given by
eq.~(\ref{Vofalp}). This integral is divergent in the IR and converges
slowly in the UV. One can (define and) evaluate it efficiently as follows:
To improve the convergence in the UV, add and subtract the Richardson
potential with suitably chosen $\Lam_R$. This potential has a quickly
convergent integral representation~\cite{Rich}. Then cancel the $q^{-2}$
singularity of the new integrand by subtracting a term with
$\alp(q) \propto q^{-2}$ and add it on again in the analytically continued
form of eq.~(\ref{varFT}).
In the end add a suitable constant to match $V(r)$ to $\Var$.

\subsection{Fitting Procedure}

As it stands, we would perform fits of the MC data to the
ansatz~(\ref{Vaofalp}) by varying the overall constant $V_0$ and the
parameters $\Lam_V$, $\sig$ and $c$ in $\aq$.
However, it turns out that $\Lam_V$ is strongly correlated with $c$
(and $V_0$), leading to a large error in $\Lam_V$ without a
corresponding large error in $\aq$. Clearly, the $\Lam$-parameter is a
bad parameterization of the physics. It is a much better idea to use
$\aqs$, where $\qs$ is {\it some} UV scale, as independent fit
parameter.  For easy comparison with the Lepage-Mackenzie method of
estimating $\aqs$ (see sect.~4.2) we choose $a \qs = 3.4018$.

We will therefore fit $V_0$, $\aqs$, $\sig$ and $c$. At present all our
fits are uncorrelated, \ie~we do not take into account correlations
between the potential data at different~$\r$
% Experience with Coulomb $+$ Linear fits seems to show~\cite{} that
% these correlations hardly affect the fit parameters or their errors.
% These fits however leave out the small $r$ data. So we can not
% exclude that the correlations have an effect at small $r$. We leave
% exploring this issue to future studies.
(we will comment on this in sect.~4.1.2).  Instead, we form the usual, naive
$\chi^2$ {}from our ansatz and the Monte Carlo data, which are
available to us in the form $V_{{\rm MC}}(\r) \pm \delta V_{{\rm
MC}}(\r)$.  To find the minimum of $\chi^2$ we use Powell's
method~\cite{NR}, which does not require knowledge of any derivatives.

\subsection{Error Analysis}

We will find that the optimal value of $c$ is always quite small, with
an error that can be of the same order as the average value of $c$.
This is unfortunate, since, we recall, for $\aq$ to be consistent with
full 2-loop evolution, any $c > 0$, but not $c =0$ is allowed.
The optimal value is somewhat irrelevant, though, because the
distribution of $c$ is often very non-gaussian, with a long tail
towards larger values of $c$.

We will also find that $c$ can be quite strongly
correlated with $V_0$ and $\aqs$, whose distribution is then also
skewed.
The reason for this correlation is that, in particular for coarse
lattices, only the first few points really ``see'' perturbative QCD,
and we then have an over-parameterization of the UV and intermediate
regime in terms of $V_0$, $\aqs$ and $c$.
As one would expect, these problems tend to go away as the data become
more precise and the lattice spacings smaller.

The correlation between the parameters $\aqs$, $c$ and $\sig$ is
usually such that they tend to move up and down
together.
The distribution of $\aqs$ and $c$ can usefully,
if somewhat simplified, be visualized as a narrow and relatively flat
ridge, whose maximum is close to the end of the ridge where both
parameters are small.  In such a situation, error estimates based on
the covariance matrix at the optimal parameter values are very
misleading.\footnote{It does not help, by the way, to consider $\ln c$
instead of $c$ as fit parameter. Note, first of all, that this would
correspond to a different assumption about the prior distribution of
$c$ (in the sense of Bayes' theorem) --- now all values of $\ln c$,
not $c$, are considered equally likely {\it a priori}. In this case
the distribution of $\ln c$ is strongly skewed in the opposite
direction, with a long tail towards negative values of $\ln c$ and a
sharp fall-off in the positive direction. In some cases the
distribution even seems to become unnormalizable, since the tail to
the left does not vanish sufficiently fast. We therefore stick to
fitting $c$, using the $\ln c$ fits as one measure of systematic
errors.}

To obtain reliable error estimates in all cases we have decided to use
the following procedure.
% \footnote{In the long run one should perform
% fits on bootstrap copies of the potential data to estimate the
% errors.}
Once the optimal values of the fit parameters are known, we
{\it sample} the total distribution around these values by varying the
fit parameters, accumulating {\it all} quantities of interest by
weighing each sample volume $dV$ with a relative probability $dV
\exp(-\chi^2/2)$. [Confer the maximum likelihood justification of the
$\chi^2$ method.] For the fit parameters themselves the number of
distinct values sampled was only $10 - 20$, so to estimate the 1-sigma
band and similar statistical measures {}from the corresponding
histograms we suitably interpolated the distribution between these
values. For other quantities, like $\Lam_V$ or $r_0$
(cf.~sect.~5.2), the number of different values accumulated is very
large and no interpolation is necessary.  We have checked with various
toy examples that this procedure gives reliable error estimates.
In the long run one should of course perform the error analysis by
bootstrapping the potential data --- an option we do not have at this
point.

For the error analysis
it would of course be nice if we could replace $c$ by a more well-behaved
fit parameter for the intermediate regime, namely one that
is less correlated  with $\aqs$ and $V_0$ and whose distribution is
more gaussian. We have tried various
options --- \eg~replacing $c$ by $q_0$ or $q_0/\sqrt{\sig}$ of sect.~5,
or by $\alp_V(\qs/3)$ --- but with the present data all have problems of
one sort or another in at least one case.
% of the theories considered.
It is possible though, we think, that some of these problems will not
exist for more accurate potential data.\footnote{For example, if one
uses $q_0$,
the problem is that the mapping {}from $\aqs$ and $q_0$ to $\Lam_V$ and
$c$ becomes nearly degenerate when $c$ is very small. The fact that
in some cases the optimal value of $c$ is {\it very}
small, might be an artifact of present data
that will disappear with better data (more precisely, once the
effective potentials have been extrapolated to large euclidean times,
 cf.~sect.~4).}

The above concerns the statistical errors. Since systematic errors
affect different quantities to different extents, they will be
discussed in the next two sections where we present explicit results for
various quantities.

\section{Results and Comparison with Other Methods}

\subsection{Overview}

We applied our scheme to potential data obtained by Monte Carlo
simulations of Wilson loops in the following lattice gauge theories:

\begin{enumerate}

\item[$\bullet$]
Pure SU(3) at $\beta \seq 6.0, 6.4$~\cite{BS} on a
$32^4$ lattice  and at $\beta \seq 6.8$~\cite{GBpriv,BSnew} on a
$48^3 \times 64$ lattice.
Potential data exist, basically, at all lattice points
that are multiples of $(1,0,0)$, $(1,1,0)$, $(1,1,1)$, $(2,1,0)$,
$(2,1,1)$ or $(2,2,1)$, and have no component larger than $16$ for
$\beta\seq 6.0$ and~$6.4$,  and no component larger than $24$ for
$\beta \seq 6.8$.
For orientation we jump ahead and remark that
the lattice spacings of these theories are between roughly $0.10$~fm and
$0.038$~fm.
% The $\beta$=$6.8$ used are somewhat
% preliminary, a more detailed analysis of the final data will appear
% in~\cite{BKS}.
We should mention that the $\beta\seq 6.8$ data
{}from~\cite{BS}, which we had first analyzed, turned out to suffer
{}from large finite-size effects --- even though they are nicely
linear at large $r$ ---
as the comparison with the new data shows.
We will see later that the new $\beta \seq 6.8$ data also show signs
(much smaller though) of finite-size and/or other systematic errors.
% since the string tension is not stable within its
% rather small errors, when data points at large $r$ are left out.
% We also observe that, compared to other theories, the fit parameters are
% not as stable when data points at short distances are deleted.
Finally, we should point out that the $\beta \seq 6.8$
data used in~\cite{myLATBi} are {}from the same Wilson loop data, but
used a preliminary extraction~\cite{GBpriv} of the potential {}from
these Wilson loops.
New SU(3) data will be analyzed with our method in~\cite{BKS}.

\vskip 1mm
\item[$\bullet$]
SU(3) with two flavors of of dynamical staggered fermions
of mass $a m\! =\! 0.01$ at $\beta \seq 5.6$~\cite{HelNftwo},  and
two Wilson fermions
($\kappa\!=\! 0.1675$) at $\beta \seq 5.3$~\cite{HelNftwo}.
The lattice size is $16^3 \times 32$ in both cases, with a spacing of
about $0.10$~fm and $0.14$~fm, respectively. The potential was calculated
at all points in a plane with no component larger than $8$.
There was no sign of string breaking  for the $r$-values considered.
It seems that the main effect of the non-zero fermion masses is to
renormalize the fit parameters. [We will see however that the
effect of unquenching can {\it not} be absorbed in a ``$\beta$-shift''
with respect to quenched QCD.]

\vskip 1mm
\item[$\bullet$]
Pure SU(2) at $\beta \seq 2.85$~\cite{UKsutwo} on a $48^3 \times 56$
lattice.  Potential  data\footnote{As
in the references above, for the SU(2) case and those with
dynamical fermions we actually used data for the ``effective
potential'' $V_T(\r)$ obtained {}from Wilson loops at euclidean times
$T$ and $T+1$, with $T\seq 3$ for the Wilson fermion case, and
$T\seq 4$ for the SU(2) and staggered fermion cases. The actual
potentials, obtained by a large $T$ extrapolation, will differ
somewhat {}from  these effective potentials.}
were available at all points with no component larger than $3$ and at
all even on-axis points $r/a=4,6,8,\ldots,24$. The lattice spacing is
about $0.028$~fm.

\end{enumerate}

\vskip 2mm

\no
Recall that $\beta$ is related to the bare lattice couplings $g_0$,
respectively $\alp_0$, by $\beta \equiv 2N/g^2_0 = N/2\pi \alp_0$.

\subsubsection{Results from the Standard Ansatz}

Using our standard ansatz of sect.~2 we could fit all data down to one
lattice spacing, with the uncorrelated $\chi^2 \approx N_{{\rm DF}}$.
In sect.~4.1.2 we will discuss in detail what possible effect the
neglect of correlations has on our results. To convince the reader
that our fits really are much better than previous ones, let us here
just note
(more details will be given as we proceed)
that in contrast to our ansatz, it is completely impossible, at least
on fine lattices, to describe the data within errors down to one, or even
several lattice spacings
with the modified Coulomb $+$ linear ans\"atze discussed in
sect.~1. This is particularly obvious from plots of the difference between
fit and data; we will present such plots below.
%For modified Coulomb $+$ linear fits one has to omit more and more
%small $r$ data points as the errors and the lattice spacings become smaller.
We also note that our fit parameters
are essentially stable with respect to variations of the fit range,
which can not be said about Coulomb $+$ linear fits on fine lattices.

{
\begin{table}[t] \centering
\begin{tabular}{ | c | l | l | l | l | l | l | c | }
\hline
Group & $n_f$ & $\beta$~ & $\aqs$ &
     ~$a \Lam_V$ & ~$a^2 \sig$ & ~~~~$c$ & $\chi^2/N_{{\rm DF}}$\\ \hline
SU(3) & 0  & 6.0  & 0.1460 & 0.1359 & 0.0484  & $3.60\cdot 10^{-4}$& 65.2/62\\
SU(3) & 0  & 6.4  & 0.1296 & 0.0986 & 0.01503 & $7.97\cdot 10^{-3}$& 89.6/66\\
SU(3) & 0  & 6.8  & 0.11512& 0.0602 & 0.00689 & $6.84\cdot 10^{-3}$&188.8/104\\
SU(3) & 0  &$6.8'$& 0.11465& 0.0586 & 0.00657 & $3.39\cdot 10^{-3}$&135.0/68~\\
SU(2) & 0 & 2.85  & 0.1671 & 0.0470 & 0.00398 & $5.01\cdot 10^{-5}$& 36.2/26\\
SU(2) & 0 &$2.85'$& 0.1678 & 0.0485 & 0.00386 & $7.53\cdot 10^{-5}$& 21.8/26\\
SU(3) & 2W & 5.3  & 0.1895 & 0.1749 & 0.1001 & $4.46\cdot 10^{-5}$ & 40.0/37\\
SU(3) & 2S & 5.6  & 0.1674 & 0.1262 & 0.0485 & $5.42\cdot 10^{-5}$ & 41.0/40\\
\hline
\end{tabular}
\vskip 1mm
\caption{Optimal fit parameters and uncorrelated $\chi^2$ for our
         standard ansatz. W, S indicates
         Wilson, respectively, staggered fermions.
         The $\beta\seq 6.8'$ and $2.85'$ data sets are explained in the
         main text. Note that $\Lam_V$ is
         {\it not} a fit parameter in our procedure; it is quoted for
         reference purposes.}
\label{OptPars}
\vskip 4mm
\end{table}

% \vskip 3mm
\begin{table}[thb] \centering
\begin{tabular}{ | c | l | l | l | l | l | l | }
\hline
Group & $n_f$ & $\beta$~ & ~~$\aqs$ &
         ~~~$a \Lam_V$ & ~~~$a^2 \sig$ & ~~~~$c$\\ \hline
SU(3) & 0  & 6.0  & 0.1470(18) & 0.140(18)  & 0.0486(5)   & 0.0011(9)\\
%SU(3) & 0  & 6.0  & 0.1470(17) & 0.143(12)  & 0.0486(5)   & 0.0010(8)\\
SU(3) & 0  & 6.4  & 0.1299(7)  & 0.0997(28) & 0.01513(25) & 0.012(6)\\
SU(3) & 0  & 6.8  & 0.11514(12)& 0.0603(4)  & 0.00690(5)  & 0.0070(8)\\
SU(3) & 0  &$6.8'$& 0.11468(17)& 0.0587(5)  & 0.00658(7)  & 0.0036(7)\\
%SU(2) & 0 & 2.85s & 0.1675(5)  & 0.0484(19) & 0.00398(8)  & 0.000129(92)\\
 SU(2) & 0 & 2.85  & 0.1675(5)  & 0.0484(19) & 0.00398(8)  & 0.00013(9)\\
 SU(2) & 0 &$2.85'$& 0.1685(8)  & 0.0510(25) & 0.00387(9)  & 0.00045(37)\\
SU(3) & 2W & 5.3  & 0.1921(22) & 0.207(20)  & 0.1002(6)   & 0.0011(10)\\
SU(3) & 2S & 5.6  & 0.1686(11) & 0.138(10)  & 0.0485(3)   & 0.00042(36)\\
\hline
\end{tabular}
\vskip 1mm
\caption{Average fit parameters and 1-sigma band for our standard ansatz,
         obtained with the error analysis procedure of sect.~3.3.}
\label{Pars}
\vskip 1mm
\end{table}
}

Our results for the optimal parameters and $\chi^2$ of the fits are
given in table~1. As already mentioned in sect.~3.3, the distribution of
$c$ is quite non-gaussian, and to obtain reliable errors we use the
procedure described there. The results are shown in table~\ref{Pars}.
As it turns out that the 1-sigma band (determined by cutting off
16\% of the distribution on each side) is typically rather symmetric
around the average of a random variable, even if the distribution itself
is quite skewed, we saw no need to quote asymmetric errors.
[In some cases we slightly shifted the mean, to put it in the middle of
the 1-sigma band.]
 Note however that the optimal parameters of table~\ref{OptPars}
usually lie very asymmetrically within the 1-sigma band quoted in
table~\ref{Pars}, for $\aqs$, $\Lam_V$ and $c$ sometimes even outside
it!

Note that in the above tables we also included data sets denoted by
$\beta\seq 6.8'$ and $2.85'$.
This was done to illustrate the effect that uncertainties
of the Monte Carlo potential data have
on our fit parameters. For the $\beta\seq 6.8'$ SU(3) data set we
omitted data points at large distances, namely all points
with at least one  component larger than $16a$.
This gives us an idea if there are finite-size (or other
large $r$) artifacts. These seem to exist, since the
string tension {}from these two fits are not consistent.
These effects are probably not quite as
large as table~\ref{Pars} indicates, since
% the
% difference in the string tensions decreases somewhat if we leave out
% the first few small $r$ points {}from the fits.
by incorporating 3-loop effects we will later find better fits,
where the difference is less significant. But the difference is still
larger than in all other cases, where the string tension is completely
stable when varying the fit range over comparable scales (in physical
units).

For the $\beta\seq 2.85'$ SU(2) set, we increased the errors of the
potential at the
first two lattice points {}from $1\cdot 10^{-4}$ to $2\cdot 10^{-4}$.
{}From the tables in ref.~\cite{UKsutwo} the extrapolation of the
effective potential to large times might easily change the values by
this amount, and, in contrast to changes at large $r$, that presumably
can be absorbed into a change of the string tension without otherwise
affecting the fit, it is not obvious that an analogous statement
will be true for short distances, where errors are much smaller.
In this case, increasing the errors at small $r$ significantly reduces
$\chi^2$, but does not lead to much different values for the fit
parameters.
We regard the two ``primed'' data sets as somewhat more reliable than
the corresponding ``unprimed'' ones.

\vskip 1mm

One might wonder why the fractional error of $\aqs$, although quite
small, is much larger than that of the lattice potential at $r\seq a$,
which in the worst case is a few times $10^{-4}$. The main reason for
this is that $\aqs$ is strongly correlated with the overall constant
$V_0$ of the potential; an increase, say, of $V_0$ can be partially
compensated by a decrease of $\aqs$ (and smaller changes of $c$ and
$\sig$). If the potential data at small $r$ are correlated, fits
that take this into account should give a more precise $\aqs$
(see below).

Note the relatively large statistical error of $\Lam_V$ in
table~\ref{Pars}, to which, as we will see below, we have to add
a systematic error that in some cases is even larger.
This large error is not
just due to the fact that
only the logarithm of $\Lam_V$ enters in $\aq$. Instead, the main
contribution to its error comes {}from the correlation
with $c$. Since both parameters are, roughly speaking, intermediate
range parameters, one should not be surprised that their
correlation is such that it leads to a much
smaller error in the running coupling than expected {}from naive error
propagation.  This was the reason
 we chose $\aqs$, not $\Lam_V$,
as the fit parameter for the UV region.

\vskip 1mm

 From table~\ref{Pars} we see
 that the quenched $\beta\seq 6.0$ and the unquenched
$\beta\seq 5.6$ SU(3) theory have almost the same string tension.
In previous studies it has often been found empirically that the
effect of unquenching can be absorbed into such a change of $\beta$.
Of course, strictly this can not be true, since it would contradict
QCD at short distances. And indeed, we find in table~\ref{Pars}
 that these two
theories have very different UV couplings $\aqs$.

\subsubsection{Correlated versus Uncorrelated Fits, Systematic Errors}

We should comment on our use of naive, uncorrelated fits
(cf.~sect.~3), that ignore any correlations the potential data might
have between different $\r$-values.  Of course, one should eventually
take these into account, preferably by bootstrapping. We do not have
this option at present, but would like to dispel any worries one might
have, that we are, say, underestimating the errors because of
this.

To investigate this question we have performed fits of all potential
data to modified Coulomb $+$ linear ans\"atze (cf.~sect.~1) using the
uncorrelated $\chi^2$, and compared them to previous results from
correlated fits with the same ansatz and fit
range~\cite{UKsutwo,BS,HelNftwo,GBpriv,Helpriv}. We estimate the
errors from the covariance matrix, which for these fits should be
reasonably accurate, since in contrast to the crossover parameter $c$
in our ansatz, for example, none of the parameters here is expected to
be strongly non-gaussian ({\it a posteriori} this is also justified by
the excellent agreement with errors from bootstrapping for weakly
correlated data).  We find:

\begin{enumerate}

\item[(i)] The uncorrelated $\chi^2$ is smaller than the correlated
one; slightly if the data are weakly correlated, {\it much} smaller
if the data have large correlations.

\item[(ii)] The central values of the fit parameters are largely
unaffected, even if the correlations are strong.

\item[(iii)] The errors from uncorrelated fits tend to be too
{\it large}, significantly so if the data are strongly correlated.

\end{enumerate}

\no
Basically, these features are well
known~\cite{GBpriv,Helpriv,corrfits}.
% \footnote{We would like to
% thank Gunnar Bali,
% Urs Heller and Rainer Sommer for discussions on these issues.}
In fact, thinking about how $\chi^2$ is affected by
(positive) correlations, and how this in turn affects the ``$\Delta
\chi^2 = 1$'' method of estimating errors, the above features are
more or less obvious.  The first point, that our fits will underestimate
$\chi^2$ if the data are correlated, is of no big concern. Our aim in
this paper is (a) to show that our ansatz for the potential is much
better than Coulomb $+$ linear ans\"atze, and (b) give an estimate of
the errors of the running
coupling extracted from our fits. If point (a) is not clear already on
purely theoretical grounds --- assuming QCD is correct --- then one
only needs to look at some pictures of the fractional difference
between our fit and the data (see below) and compare them with
corresponding pictures from Coulomb $+$ linear fits
(see~\eg~\cite{BS}).

For the cases at hand, the only potential data that have strong
correlations are those for pure SU(2).
In this case, our errors for
the parameters of $\aq$ are likely to be over- not
underestimates, according to (iii).
[In cases where the first few $r$-values had to be omitted from the
Coulomb $+$ linear fits, we
can not exclude that the potential data have
non-negligible correlations among these points.
If so, correlated fits with our ansatz would allow us
to determine $\aqs$ more precisely.]

We add some remarks from our experience of comparing (uncorrelated)
fits to our ansatz and to modified Coulomb $+$ linear ans\"atze.
We found that it is somewhat of an art to extract
parameters from Coulomb $+$ linear fits, in particular on finer
lattices. Even with the lattice artifacts taken into account as
described in sect.~1, and the Coulomb charge $e$ replaced by the
mock running charge $e-f/r$, one has to omit many small $r$ points
from the fit, and it is often not so clear if one ever reaches a
plateau at which the parameters stabilize. For the $\beta\seq 6.8$
SU(3) data, for instance, neither $e$, $f$, or the coefficient
of the lattice correction seems to stabilize; $f$ even becomes
negative at some point. One clearly sees that this type of ansatz
does not incorporate the correct physics.

Out fits, in contrast, are stable with respect to varying the fit
range.\footnote{There is a slight but noticeable dependence on the fit
range for the $\beta\seq 6.8$ SU(3) data, where (cf.~the previous
remarks on the $\beta\seq 6.8'$ data set, and below) it is just as
likely to be due to the data, rather than our ansatz.}  Furthermore,
we can at least argue --- and we will in sect.~4.2 --- that our UV
parameter $\aqs$ is most accurately determined at the shortest
distances, but no such argument is possible for the unphysical
``running charge'' $e-f/r$.

\vskip 1mm

We now turn to the systematic errors of $\aqs$, $\sig$ and $c$
in our ansatz. There
are two potential sources of such errors: (i) our ansatz for the
continuum running coupling, and (ii) remaining lattice artifacts not
taken into account in our ansatz for the lattice potential in terms of
the continuum coupling, eq.~(\ref{Vaofalp}).  Of course, in general
we can not rigorously disentangle and separately investigate these
two effects, since our fit necessarily involves both.
[We can disentangle them at the 1-loop level, as we will discuss
in sect.~4.3.]
It is pretty clear, however,
that lattice effects will have a very small direct influence on $\sig$
and probably also on $c$. They might have an effect on $\aqs$, and
through the correlations of the fit parameters this would then {\it
indirectly} influence $c$ and to a smaller extent also $\sig$.

The question whether $\aqs$ has errors due to lattice artifacts is
sufficiently important to be discussed separately in the next
subsection.  Here we will discuss the errors of the fit parameters due
to our ansatz for the coupling.
% Modulo perhaps some slight signs of
% systematic errors in $\aqs$, to be discussed below, our
     Our  fits in
themselves give no indication of significant systematic errors in this
ansatz.
% We already mentioned that the $\beta \seq 6.8$ data are
% somewhat exceptional in that the fit parameters are not completely
% stable when leaving out small or large data points.
% (the same is true
% if one uses Coulomb $+$ linear fits~\cite{GBpriv}). At present though,
% this seems more likely to be a problem of the data than of our ansatz.
% More on this will be said later.

{
\begin{table}[t] \centering
\begin{tabular}{ | c | l | l | l | l | l | l |  c | c | }
\hline
Group & $n_f$ & $\beta$~ & $\aqs$ &
   ~$a \Lam_V$ & ~$a^2 \sig$ & ~~~$c$ & $a_1$ & $\chi^2/N_{{\rm DF}}$\\ \hline
SU(3) & 0  & 6.0  & 0.1450 & 0.1081 & 0.0488  & $2.32\cdot 10^{-4}$ & 0.5 &
      64.1/62\\
SU(3) & 0  & 6.4  & 0.1286 & 0.0876 & 0.01530 & $8.77\cdot 10^{-3}$ & 0.5 &
      88.8/66\\
SU(3) & 0  & 6.8  & 0.11440& 0.0549 & 0.00718 & $1.23\cdot 10^{-2}$ & 0.5 &
      153.5/104\\
SU(3) & 0  &$6.8'$& 0.11405& 0.0532 & 0.00686 & $7.03\cdot 10^{-3}$ & 0.5 &
      119.0/68~\\
SU(2) & 0 & 2.85  & 0.1663 & 0.0482 & 0.00387 & $3.37\cdot 10^{-6}$ & 1.5 &
      26.2/26\\
SU(2) & 0 &$2.85'$& 0.1667 & 0.0487 & 0.00379 & $3.84\cdot 10^{-6}$ & 1.5 &
      17.8/26\\
SU(3) & 2W & 5.3  & 0.1883 & 0.2021 & 0.0994 & $6.12\cdot 10^{-5}$ & 1.5 &
      35.6/37\\
SU(3) & 2S & 5.6  & 0.1664 & 0.1391 & 0.0480 & $3.01\cdot 10^{-5}$ & 1.5 &
      37.0/40\\
\hline
\end{tabular}
\vskip 1mm
\caption{Optimal fit parameters and $\chi^2$ for $a_1 \neq 1$ ans\"atze.}
\label{OptParsa}
\vskip 3mm
\end{table}

% \vskip 3mm

\begin{table}[t] \centering
\begin{tabular}{ | c | l | l | l | l | l | l | c | }
\hline
Group & $n_f$ & $\beta$~ & ~~$\aqs$ &
         ~~~$a \Lam_V$ & ~~~$a^2 \sig$ & ~~~~$c$ & $a_1$\\ \hline
%
%SU(3) & 0  & 6.0  & 0.1462(15) & 0.118(12)  & 0.0492(7)   & 0.00100(81) &
%%0.5\\
SU(3) & 0  & 6.0  & 0.1462(15) & 0.118(12)  & 0.0492(7)   & 0.0010(8) & 0.5\\
SU(3) & 0  & 6.4  & 0.1289(7)  & 0.0892(37) & 0.01545(33) & 0.014(8)    & 0.5\\
SU(3) & 0  & 6.8  & 0.11443(13)& 0.0550(5)  & 0.00719(6)  & 0.0129(17)  & 0.5\\
SU(3) & 0  &$6.8'$& 0.11408(17)& 0.0533(8)  & 0.00689(10) & 0.0076(18)  & 0.5\\
SU(2) & 0  &2.85  & 0.1669(7)  & 0.0496(17) & 0.00385(9)  & 0.000012(9)& 1.5\\
SU(2) & 0  &$2.85'$&0.1682(11) & 0.0521(27) & 0.00376(10) & 0.000042(37)& 1.5\\
%SU(3) & 2W& 5.3  & 0.1916(27) & 0.225(18)  & 0.0993(6)   & 0.00071(62) & 1.5\\
SU(3) & 2W & 5.3  & 0.1916(27) & 0.225(18)  & 0.0993(6)   & 0.0007(6) & 1.5\\
SU(3) & 2S & 5.6  & 0.1685(16) & 0.152(9)   & 0.0479(3)   & 0.00022(18) & 1.5\\
\hline
\end{tabular}
\vskip 1mm
\caption{Average fit parameters and 1-sigma band for $a_1 \neq 1$
         ans\"atze.}
\label{Parsa}
\vskip 1mm
\end{table}
}

The  only real check of our ansatz for $\aq$, then,      is to try
to find other ans\"atze that give similarly good or better
fits. This is not so easy. {}From sect.~2 we know how to
incorporate higher loops into $\aq$. However, each loop adds at least
two new parameters (one corresponding to the coefficient in the
$\beta$-function, the other is a non-perturbative parameter arising
{}from our method of avoiding the Landau pole), and this often leads to
stability problems with the fits, since the limited part of the
intermediate and UV region available in practice has effectively been
vastly overparameterized. Another way of saying this, is that with
present data  non-perturbative and all-order effects
apparently become important before we can clearly discern 3-loop
effects.

We have tried various ans\"atze that fully or partially take into
account 3- and 4-loop effects. Even if they give stable results, most
give significantly better fits only for one of the theories we are
considering.  This might simply reflect a (not even large) fluke of
the data in question.  The overall best method we found has one
new parameter $a_1$, that takes into account the leading 3-loop
effects.  It corresponds to the choice
$\lam(t) = \ln^b(c_1 + c~\e^{a_1 t})$ in eq.~(\ref{myalp}).
Even these fits are not always
stable and so in tables~\ref{OptParsa} and~\ref{Parsa} we show results
for suitably chosen fixed
$a_1$. The values of $\chi^2$ are significantly better for the last
four theories shown.  None of the other ``sporadically'' better fits
we found seems to lead to a much larger variation in fit parameters
(or other quantities obtained {}from $\aq$, cf.~sect.~5) than that
observed when comparing tables~\ref{OptPars} and~\ref{Pars}
with tables~\ref{OptParsa} and~\ref{Parsa}.

Comparing tables~\ref{Pars} and~\ref{Parsa} we see that the systematic
errors of $\aqs$ and $\sig$ are never much larger, if larger at all,
than their statistical errors. It is significantly larger in some
cases for $\Lam_V$.
Since $c$ is really a different parameter for different $a_1$, it is
not surprising that it can also change significantly.
More relevant, however, is the question of
how much $\aq$ {}from different fits differ in the intermediate
region. We find, not surprisingly, that in this region $\aq$ is
typically even better determined than at its ``edges''.
% In other
% words, the overall fractional error of $\aq$ decreases as one goes
% towards the intermediate region {}from either the UV or IR edge, where
% the fractional error of $\aq$ is equal to that of $\aqs$ and
% $\sig$, respectively.

\begin{figure}
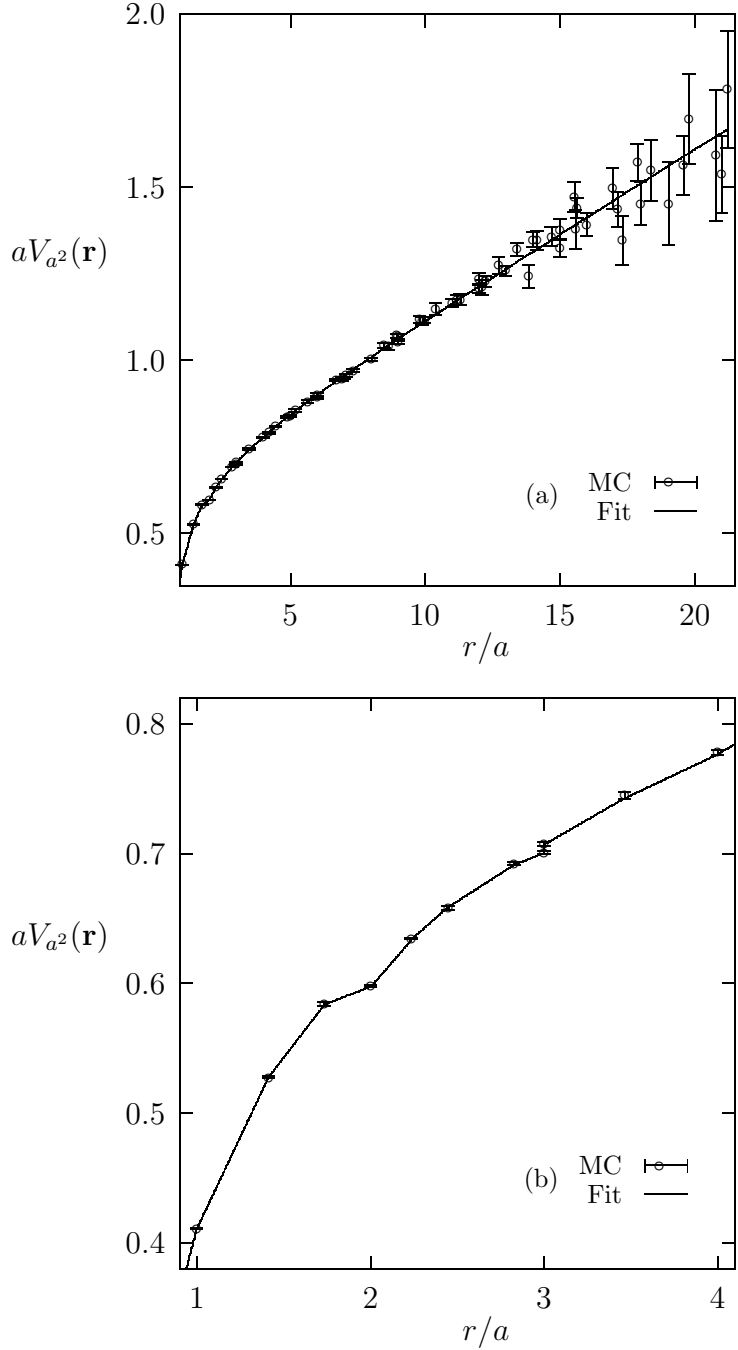
 \centering
%% \include{../../c++/fgrid/gnuplot/Vb6_2_0.001}
% GNUPLOT: LaTeX picture
\setlength{\unitlength}{0.240900pt}
\ifx\plotpoint\undefined\newsavebox{\plotpoint}\fi
\sbox{\plotpoint}{\rule[-0.175pt]{0.350pt}{0.350pt}}%
%
\caption{
Comparison of Monte Carlo data (circles with error bars) for the
lattice potential in pure SU(3) gauge theory at $\beta\seq 6.0$
with $\Var$ {}from our standard fit (lines), calculated on
the lattice points and connected with straight lines to guide the eye.
Shown is an overview~(a) and the small~$r$ region~(b).}
\end{figure}

\begin{figure}
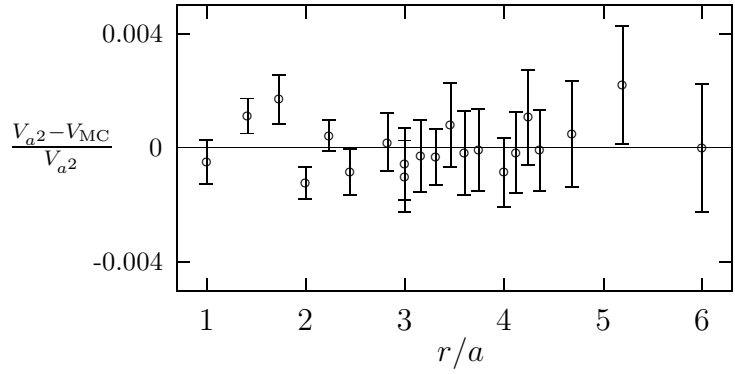
 \centering
%%\include{../../../c++/fgrid/gnuplot/fdiff.su2_b1.5_c30_N2}
% GNUPLOT: LaTeX picture
\setlength{\unitlength}{0.240900pt}
\ifx\plotpoint\undefined\newsavebox{\plotpoint}\fi
%
\caption{Fractional difference between our $a_1 \seq 1.5$ fit and the
         Monte Carlo results for the $\beta\seq 2.85'$ pure SU(2)
         potential in the small $r$ region.}
\end{figure}

\begin{figure}
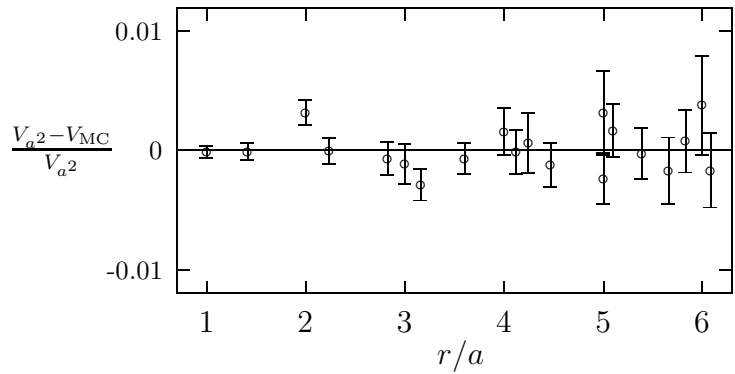
 \centering
%%\include{../../../c++/fgrid/gnuplot/fdiff.Nf2S_b1.5_c30_N2}
% GNUPLOT: LaTeX picture
\setlength{\unitlength}{0.240900pt}
\ifx\plotpoint\undefined\newsavebox{\plotpoint}\fi
%
\caption{Same as figure~2, for our $a_1 \seq 1.5$ fit of the potential
         of $\beta\seq 5.6$ SU(3) with two staggered fermions.}
\end{figure}

\vskip 2mm

For illustration, one of our fits is shown in figure~1. One sees
that every little bump and wiggle of the data is reproduced by our fit.
Since the errors are too small to be resolved in graphs of the potential
itself, we show in figures~2 and~3 examples of the fractional difference
between our fit and the Monte Carlo data.

Though these plots do not
in any obvious way indicate that there are lattice artifacts not
reproduced by our ansatz,\footnote{In the process of searching for
possible systematic
errors due to lattice artifacts, we have also checked what
happens if we use $\aq$ instead of $\alpha_V(\qh)$ in eq.~(\ref{Vaofalp}).
It turns out that despite the slow running of the coupling this
{\it does} have a significant effect: It is not possible to get good fits
at small $r$; the points $r/a \seq \sqrt{2}, \sqrt{3}, 2$ are impossible to
fit simultaneously. It is of course not too surprising that one should use
$\alp_V(\qh)$ in~(\ref{Vaofalp}), since the integrand of $\Var$ should be
periodic in all $q_i$.}
one could take the fact that the small $r$ region contributes a
relatively larger fraction to the total $\chi^2$ as a hint of such
a systematic error.  However, this is at present just as likely to
come {}from slight underestimates of the error or shifts in the central
values of the potential data (compared to the true values) than {}from
systematic errors in our ansatz. We recall that obtaining the true
potential involves an extrapolation to large euclidean times.
Uncertainties in the data arise, because either this extrapolation
is not done and one simply quotes an effective potential from a fixed
time, or, if the extrapolation is done, it is hard to independently
assess its reliability for small $r$ (where the quoted errors
are very small), since there was no theoretical expectation of what
exactly the answer  should be.  This is underscored by the fact that
slightly different procedures of extracting the potential {}from Wilson
loops tend to lead to correlated changes of the potential. This can
have a large effect on the $\chi^2$ of our fit, and lead to
significant changes of the fit parameters.\footnote{We know
for a fact that this is true for the existing
$\beta\seq 6.8$ pure SU(3) data,
where Gunnar Bali generously provided us with potentials extracted
by slightly different procedures {}from the Wilson loop data.
%Of course, this theory is an extreme case, in that it is very difficult
%to simulate, since very big lattices are required to reach the
%confinement regime. It is perhaps the limit of what can reasonably
%be done with unimproved pure gauge actions.
}

\vskip 2mm

{}From tables~\ref{OptPars} to~\ref{Parsa} it seems
the theories under consideration fall into two classes:
Pure SU(3), and the rest. For the former better
fits are obtained for $a_1 \! < \! 1$ instead of $a_1 \seq 1$, whereas
for the latter $a_1 \! > \! 1$ is preferable. Presumably related
is the fact that in the pure SU(3) cases $c$ is consistently
larger than for the other theories.

The optimistic view would be that this is due to the size of the
third $\beta$-function coefficient. It is not known in the V scheme,
but if the $\MSbar$ scheme is any guide, this can certainly not be
sole explanation, since in this scheme~\cite{threeloop}
$a_1 \simeq 0.600$ is the
same for pure SU(2) and SU(3), and $a_1 \simeq 0.625$ for SU(3) with
$n_f \seq 2$.

In certain respects the pure SU(3) data seem not to behave as
 ``nice'' as the others. For
example, the $\chi^2$ of the fits are not as good, and various quantities
are less stable with respect to changing the form of our ansatz for
$\aq$ or when leaving out small $r$ data points (cf.~also sect.~5).
The latter is true even though
for pure SU(3) the decrease of $\chi^2$ when allowing  $a_1 \neq 1$ is
relatively smaller than for the other theories. One reason that might
at least partially explain the relative stability and other ``nice''
features of the SU(2) data, as compared with the $\beta \seq 6.8$
data which naively would be expected to behave quite similar, is that at
large $r$ the SU(2) data
consist only of on-axis points. Another reason could
be that the SU(2) data have significant correlations in $r$, at least for
large distances~\cite{UKsutwo}.  However, this does not
distinguish the $n_f \seq 2$ theories {}from the other pure SU(3) cases.

One fact that does differentiate the two classes is that only
for the pure SU(3) data
was the potential extrapolated to large euclidean times;
in the other cases only effective potentials were available.
Another point to note is that for the pure SU(3) theories
potential data exist for much larger
distances.
%% , where the signal to noise ratio is worse.
We performed a
fit of the $\beta\seq 6.0$ data where we only included points that also
exist for unquenched SU(3) at $\beta\seq 5.6$. The value of $c$
decreases significantly, and, more generally, the SU(3) data behave
more similar to those of the second class also in other (though not all)
respects.

Although we do not understand why $c$ should decrease,
the facts mentioned in the last paragraph would seem to provide
the best guess, at present, of why the
theories under consideration should appear to fall in two
classes.\footnote{To be sure, note that the ``smoother'' behavior
of the second class
is not due to smaller errors. Despite the extrapolation,
it is the pure SU(3) data that have much smaller errors.}
The ``nicer'' behavior of the second class might disappear once
extrapolated to large euclidean times. In fact, it was observed
in~\cite{UKsutwo} that the difference between the (on-axis)
effective SU(2) potential and the extrapolated one is not smooth in $r$.
If this is the problem,
one should investigate if it is possible to perform the extrapolation
in a smoother way.

Note that in itself the small value of $c$ in the second class
of theories is somewhat questionable, since $c\seq 0$ is not
consistent with full 2-loop evolution.
On the one hand, it is only $\ln c$ that enters into the UV expansion,
so a small value of $c$ does not necessarily violate the
principle of naturalness. However, note that the value of $c$
is even smaller for the $a_1 > 1$ fits than for the $a_1 \seq 1$ ones.
This makes a certain
mathematical sense, but one might question it on physical grounds:
Recall that the $\chi^2$ of these fits is significantly better than
those with $a_1 \seq 1$. If we take this to mean that we are
really seeing 3-loop effects, then it is surprising that at the same
time the 2-loop effects are suppressed due to the small value of
$c$. However, since a small $c$ always suppresses  both 2-loop {\it
and} (the leading) 3-loop effects, this argument is not
conclusive. Only new data, with careful consideration of finite-size
and extrapolation effects, will help to clarify these questions.

%% \vskip 2mm

\subsubsection{Estimate of the String Tension}

For future reference we would like combine our results and
quote estimates of the string tension for the various theories we
studied. To obtain the best values we should not just take into
account the results in tables~\ref{Pars} and~\ref{Parsa}, but also
the following: We are performing a global fit and do not want any
potential uncertainties at small $r$ to distort the fit at large $r$.
Therefore we fix $\aqs$ at its average value obtained {}from a fit with all
data points, and leave out successive short-distance points until
$\sig$ stabilizes within errors. In most cases this leads to no
significant change.  Combining all
information in a semi-objective manner --- being more systematic
about this seems pointless in view of the uncertainties discussed
above --- we obtain the estimates in table~\ref{sigma}, where they
are compared with previous results {}from Coulomb $+$ linear fits.
The final results for the
%% UKQCD and Wuppertal
pure SU(2) and SU(3) ($\beta \leq 6.4$) data
are quoted in~\cite{Luthree}, respectively~\cite{Lufour};
cf.~\cite{GBpriv} for $\beta\seq 6.8$.
The $n_f\seq 2$ results are {}from~\cite{HelNftwo}.

\begin{table}[tb] \centering
\begin{tabular}{ | c | l | l | l | l | l | l | c | }
\hline
Group & $n_f$ & $\beta$~ & $a^2\sig$ (our~fit) & $a^2\sig$ (C$+$L fit)\\ \hline
SU(3) & 0  & 6.0  & ~0.0491(8)  & ~0.0513(25)\\
SU(3) & 0  & 6.4  & ~0.0153(4)  & ~0.0139(4)\\
SU(3) & 0  & 6.8  & ~0.0070(3)  & ~0.00705(14) \\
SU(2) & 0  &2.85  & ~0.00375(14)& ~0.00354(26)\\
SU(3) & 2W & 5.3  & ~0.0992(9)  & ~0.0976(9)\\
SU(3) & 2S & 5.6  & ~0.0480(5)  & ~0.0481(5)\\
\hline
\end{tabular}
\vskip 1mm
\caption{Comparison of our estimate of the string tension in lattice
         units with previous
         ones {}from (modified) Coulomb $+$ linear fits.}
\label{sigma}
\vskip 1mm
\end{table}

As expected, our results are generally in good agreement with
previous estimates, though often more precise.
 An exception is $\beta\seq 6.4$ pure SU(3),
where our result is significantly higher than the latest
Coulomb $+$ linear estimate (though it agrees much better with a
previously published estimate~\cite{BS} and our own estimate
from  uncorrelated Coulomb $+$ linear fits).

Since our method takes into account the corrections to the
asymptotic linear behavior more accurately, one would expect that
it gives better estimates of the string tension
than previously possible. That our errors
are not always smaller than previously quoted ones, is
due to the fact that we were rather careful about systematic errors.
In particular, for $\beta\seq 6.8$ the quoted error is almost
completely of systematic origin. If one were similarly careful about
systematic errors for Coulomb $+$ linear fits, their errors would be
as least as large, often much larger.

\subsection{Lattice Artifacts, $\aqs$, and the Lepage-Mackenzie Method}

Our precise estimate of $\aqs$ depends on including all points
down to $r=a$ in our fits. Leaving out $r=a$ does in some  cases
lead to significant shifts in the central value. This is illustrated
in table~\ref{Parswosmallr}, where we show results {}from
fits where data points at small $r$ have been ignored.

\begin{table} \centering
\begin{tabular}{ | c | l | l | l | l | l | l | c | c | c |}
\hline
Group & $n_f$ & $\beta$~ & $\aqs$ &  ~$a \Lam_V$ &
 ~~$a^2 \sig$ & ~~~~$c$ & $a_1$ & $r_{{\rm min}}$ & $\chi^2/N_{{\rm DF}}$\\
\hline
SU(3) & 0  & 6.0& 0.1450 & 0.1081 & 0.0488  & $2.3\cdot 10^{-4}$ & 0.5 &
                                              1         & 64.1/62\\
      &    &    & 0.154  & 0.165 & 0.0502  & $1.8\cdot 10^{-2}$ & 0.5 &
                                              $\sqrt{2}$ &  59.8/61\\
      &    &    & 0.149  & 0.139  & 0.0497  & $3.2\cdot 10^{-3}$ & 0.5 &
                                              $\sqrt{3}$ &  59.0/60\\
\hline
% SU(3) & 0  & 6.4\\
%% 6.8F:
SU(3) & 0  & 6.8 & 0.1144& 0.0549 & 0.00718 & $1.2\cdot 10^{-2}$ & 0.5 &
                                              1         & 153.5/104\\
      &    &     & 0.1163& 0.0611 & 0.00758 & $4.9\cdot 10^{-2}$ & 0.5 &
                                             $\sqrt{2}$  & 102.3/103\\
      &    &     & 0.1163& 0.0610 &0.00758 & $4.8\cdot 10^{-2}$ & 0.5 &
                                             $\sqrt{3}$ & 102.3/102\\
      &    &     & 0.1160& 0.0602 &0.00754 & $4.1\cdot 10^{-2}$ & 0.5 &
                                             2          & ~91.4/101\\
\hline
%% 2.85s:
SU(2) & 0  &2.85 & 0.1663 & 0.0482 & 0.00387 & $3.4\cdot 10^{-6}$ & 1.5 &
                                              1         &  26.2/26\\
      &    &     & 0.1677 & 0.0485 & 0.00374 & $1.2\cdot 10^{-6}$ & 1.5 &
                                             $\sqrt{2}$ &  15.6/25\\
      &    &     & 0.1667 & 0.0488 & 0.00377 & $3.8\cdot 10^{-6}$ & 1.5 &
                                             $\sqrt{3}$  &  13.8/24\\
\hline
% SU(3) & 2W & 5.3\\
SU(3) & 2S & 5.6 & 0.1664 & 0.1391 & 0.0480 & $3.0\cdot 10^{-5}$ & 1.5 &
                                              1         &  37.0/40\\
      &    &     & 0.168  & 0.148 & 0.0477 & $6.7\cdot 10^{-5}$ & 1.5 &
                                             $\sqrt{2}$ &  35.9/39\\
%      &    &     & 0.1682 & 0.1478 & 0.0477 & $6.7\cdot 10^{-5}$ & 1.5 &
%                                             $\sqrt{2}$ &  35.9/39\\
%      &    &     & 0.1780 & 0.1541 & 0.0472 & $0$                & 1.5 &
%                                              2         &  27.4/38\\
\hline
\end{tabular}
\vskip 1mm
\caption{Optimal  fit parameters for some $a_1 \neq 1$ fits with
         small $r$ points left out. $r_{{\rm min}}$ indicates the smallest
         point included in the fit. Although not shown, the errors of
         $\aqs$ and especially $c$ rise rapidly
         as $r_{{\rm min}}$ is increased, except for the $\beta\seq 6.8$
         data.}
\label{Parswosmallr}
\vskip 1mm
\end{table}

If one had very precise data on a relatively fine lattice, leaving out
$r \seq a$ should not prevent an accurate determination of $\aqs$.
This seems to be the case for the SU(2) data, as
table~\ref{Parswosmallr} indicates.  However, if the data are not that
precise, leaving out $r\seq a$ is {\it expected} to lead to much larger
errors, for the following reason:
We will see in sect.~5 that the momentum scale $\qs$ corresponds
to a distance scale of roughly
$r^\star \simeq 1.53/\qs \simeq 0.45 a$. So, ignoring
$r\seq a$ will make it much harder to obtain precise information
at scale $r^\star$.

One therefore should not be surprised that the determination of $\aqs$
rapidly deteriorates as one leaves out data points on the coarser
lattices.  A bit surprising is the fact
that for the $\beta\seq 6.8$ SU(3) case the value of $\aqs$ {}from a
full fit seems incompatible (on the $3-4$ sigma level)
with that {}from fits without $r\seq a$,
which still have very small errors and a much better $\chi^2$
(for the $\beta\seq 6.8$ results shown in table~\ref{Parswosmallr} the
error of $\aqs$ is always less than $4 \cdot 10^{-4}$).

Since it is not obvious that our ansatz~(\ref{Vaofalp}) really
incorporates lattice artifacts with sufficient accuracy at one lattice
spacing, the question arises whether our method indeed breaks down at
$r\seq a$ in this case (then the value of $\aqs$ {}from
fits ignoring $r\seq a$ would be more accurate),
or if instead the value of $\aqs$ {}from the full fit is more
reliable, and the change when leaving out the first point is just the
kind of ``edge effect'' that one might expect if one has data with very
small
(perhaps slightly underestimated)
errors, and considers a quantity that depends mainly on
%% one point at
the ``boundary'' of the data set.

Although this question can not really be answered with
certainty at present, we would like to argue that the second
possibility is more likely, \ie~the value of $\aqs$ {}from the full fit
is quite reliable also for SU(3) at $\beta\seq 6.8$.
We already mentioned  that there are some uncertainties in the
$\beta\seq 6.8$ data at large $r$, so the same might be true at small
$r$ (where previously no independent check of the data was
possible), in particular in view of the fact that the
pure SU(2) data do not seem to have this problem.
%% cf.~tables~\ref{Parswosmallr}, \ref{Pars} and~\ref{Parsa}.
The lattice
spacing in this case is comparable to that of the SU(3) theory,
even smaller, and the errors at short distances are about as good.
[At $r\seq a, \sqrt{2} a$ the potential in the SU(2) case has relative
errors of $4\cdot 10^{-4}$, $3\cdot 10^{-4}$, respectively, whereas
for SU(3) the corresponding numbers are
$2\cdot 10^{-4}$, $4\cdot 10^{-4}$.]

\omitThis{
Even if we get a good fit including $r=a$, this could be due to the
compensation of one error by another, \ie~the value of $\aqs$ {}from our
fit is not obviously correct because the fit is
good.\footnote{However, as lattice spacings get smaller and the data
more precise, the goodness of the fit in itself becomes a stronger and
stronger indication that the fit result for $\aqs$ is also correct
within its error.}
}

\vskip 1mm

These arguments would be more convincing if we had a completely
independent estimate of $\aqs$ that we could compare with.
Fortunately, such an estimate exists.  Namely,
Lepage and Mackenzie~\cite{LM} have argued that the running coupling
$\aq$ at some UV scale can be determined by comparing the
non-perturbative expectation value of the plaquette operator,
$W_{11} \equiv \frac{1}{N} \langle {\rm Tr} ~U_{\plaq} \rangle$,
as measured in the Monte Carlo simulation, with
its expansion in terms of $\aq$. {}From the results
of~\cite{HasHas,Weisz,Fischler,HelKar,Helpriv} this expansion can
be written as
 \bea\label{lnWexp}
 -\ln W_{11} =  \pi C_F \aq ~\Bigg[ 1 - \aq \Bigg(\frac{11 N}{12 \pi}
    \ln\Big(\frac{6.7117}{aq}\Big)^2 + (\delta_f +\frac{1}{6\pi}
          \ln(aq)^2 ) \ts n_f \Bigg) \Bigg] \nonumber \\
        + ~\O(\aq^3) ~,~~~~~~
 \eea
where $\delta_f \seq -0.060$ for staggered and
$\delta_f \seq -0.105$ for Wilson ($r\seq 1$) fermions.

The next question is what $q$ to choose in eq.~(\ref{lnWexp}).
Were the expansion known to large orders, one could extract $\aq$ at
basically any $q$.
% Were the rhs of this equation known to all orders it would allow one
% to determine $\aq$ at any $q$.  For any finite truncation
For a given truncation of the series, however,
there is an
``optimal scale'' $a \qs$ at which to extract $\aq$. Intuitively $\qs$
is the ``natural scale'' of the ``physical process'' in question.
Let us write the leading non-trivial contribution to a lattice
quantity as $\alp_V(q^\star)\int_{-\pi/a}^{\pi/a} d^4q ~F(q)$.
[Depending on the quantity in question this might be the 1-loop
or the tree level contribution.]  To this order the
natural definition of $\qs$ is\footnote{According to
our philosophy it would seem more natural to use $\alp_V(\hat{q})$ in
eq.~(\ref{qstardef}) to define a quantity $\hat{q}^\star$. If this is
done, one would also have to use $\hat{q}$ in eq.~(\ref{lnWexp}), for
example. However, in that equation one wants to think of $\aq$ as a
continuum object, so the use of $\hat{q}$ might not appear natural. We
will not pursue this question here, since in the context of the scale
determination the use of $q$ or $\hat{q}$ will not make much
difference in the final results.}
 \be\label{qstardef} \alp_V(\qs) ~ \int_{-\pi/a}^{\pi/a} ~d^4q ~ F(q)
 ~\equiv~ \int_{-\pi/a}^{\pi/a} ~ d^4q ~\aq ~ F(q) ~.
\ee
Using the 1-loop expansion of $\aq$ around $q\seq \qs$
leads to an explicit formula for~$\qs$,
 \be\label{qstar}
    \ln q^{\star 2} ~=~  \frac{\int_{-\pi/a}^{\pi/a} ~d^4q ~F(q) ~\ln q^2}
                            {\int_{-\pi/a}^{\pi/a}  ~d^4q ~F(q)} ~.
 \ee
The integrands $F(q)$ appropriate for planar Wilson loops can be
found, for example, in~\cite{HelKar}. For the plaquette
$W_{11}$ or $\ln W_{11}$  one gets the right $\qs$ if one simply
uses $F(q)\equiv {\rm const}$. Using the methods of sect.~3.1
we obtain $a\qs = 3.4018$.

Expressed in terms of the physical coupling $\aqs$ the expansion of
$\ln W_{11}$ is expected to be well-behaved, \ie, at least the first
few higher order
coefficients in the square bracket of eq.~(\ref{lnWexp}) are expected
to roughly follow the pattern established by the first two, which in
this case both are $\O(1)$.
% respectively, 1 and about $N/3$.
% in contrast to the expansion in the bare
% lattice coupling $\alp_L$, which has large higher order coefficients
% due to large ``tadpole'' contributions~\cite{LM}.
Using the measured $W_{11}$ to determine $\aqs$ {}from (\ref{lnWexp})
with the higher order coefficients set to $0$, we will therefore quote
it with nominal error
%% $(N/3)^2 \aqs^3$,
$\aqs^3$, which would be the actual error
%% (for small $\aqs$) if the next coefficient had magnitude $(N/3)^2$.
(for small $\aqs$) if the next coefficient had
magnitude~1.\footnote{Actually, for gauge group SU(2) the second
coefficient is smaller by a factor $2/3$ (for $n_f=0$) when compared
to SU(3), and we indeed find that various errors in the SU(2) case are
smaller by roughly a factor $(2/3)^2$.} The value of $\aqs$ obtained
in this way will be denoted by $\alp_V(\qs, W_{11})$. These numbers
are collected in table~\ref{WalpLM} for the theories under
consideration.

Note that other quantities besides $W_{11}$ can be used in the
Lepage-Mackenzie prescription (see figure~4 below). Each quantity comes
with its own $\qs$. What distinguishes $W_{11}$ is that among all
simple quantities whose $\aq$ expansion is known to second order
it has the largest $\qs$. It should therefore give the most reliable
results for the coupling in the UV.

By assuming 2-loop evolution to be accurate at $\qs$, one can
obtain an estimate of the $\Lam$-parameter {}from $\aqs$. For future
reference we quote in table~\ref{WalpLM} the value of $\Lam_V$
obtained {}from $\alp_V(\qs, W_{11})$ by assuming that $\aq$ is
given by the leading part of 2-loop evolution, $\beta_0 \aq =
1/(t+b\ln t)$. This value is denoted by $\Lam_V^{{\scsc (2)}}(W_{11})$.

\begin{table} \centering
\begin{tabular}{| c | l | l | c | l | l | }
\hline
Group & ~$n_f$ & ~$\beta$ & $W_{11}$ &
         $\alp_V(q^\star,W_{11})$ & $a \Lam_V^{{\scsc (2)}}(W_{11})$ \\ \hline
SU(3) & ~0  & 6.0  & 0.59368 & 0.1519~(35)  & 0.169~(13)  \\
SU(3) & ~0  & 6.4  & 0.63064 & 0.1302~(22)  & 0.0968~(64) \\
SU(3) & ~0  & 6.8  & 0.65922 & 0.1153~(15)  & 0.0581~(34) \\
% SU(3) & ~0  & 9.0  & 0.75614 & 0.0731~(4)   & 0.00410~(16) \\
%                                               0.004104
% SU(3) & ~0  & 18.0 & 0.88450 & 0.03040~(3) & 1.05(2)$\cdot 10^{-7}$\\
%                                              1.04749
SU(2) & ~0 & 2.85  & 0.70571 & 0.1712~(50)  & 0.0554~(74) \\
SU(3) & ~2W & 5.3  & 0.53354 & 0.1991~(79)  & 0.255~(26) \\  % 0.2551
SU(3) & ~2S & 5.6  & 0.56500 & 0.1788~(57)  & 0.185~(19) \\  % 0.184953
\hline
\end{tabular}
\vskip 3mm
\caption{The expectation value of the plaquette, the
         coupling at the UV scale $a \qs =3.4018$ determined by the
         Lepage-Mackenzie prescription, and the value of the
         $\Lam$-parameter obtained by assuming $\aq$ to be given by
         the leading part of 2-loop evolution.}
\label{WalpLM}
\vskip 3mm
\end{table}

Comparing with tables~\ref{OptPars} to~\ref{Parsa}
 we see that our and the Lepage-Mackenzie
estimate of $\aqs$ are  consistent within their errors,
our values being somewhat smaller (in particular for theories on
coarser lattices). It
would be nice to compare our and the Lepage-Mackenzie estimate in
a situation where the errors are much smaller. This is indeed
possible. In~\cite{DHLM} Wilson loops up to size $8\times 8$ were
generated for pure SU(3) on very fine $16^4$ lattices. We used their
$\beta \seq 9.0$ and 18.0 data data to extract the potential
(using times $T\seq 7$ and~8)
at the first seven on-axis points, and then fitted them to our standard
ansatz. Obviously $\sig$ can not be fitted {}from such data, so we simply
set it to some reasonably value estimated by scaling, the precise
value being irrelevant. For simplicity we also fixed $c \seq 0.001$
(again, the precise value of $c$ has hardly any effect).
We performed fits using all $r\! \geq \! a$ data, and also fits with
the first point left out. In table~\ref{alpLMbbig}
 we show our  results for $\aqs$
and the corresponding results obtained by the Lepage-Mackenzie method,
which is very accurate for these fine lattices. Note the perfect agreement
for the $r\! \geq \! a$ fits. When leaving out the first point, the
central value of the fitted $\aqs$ becomes significantly worse
--- despite the much better $\chi^2$ --- though with
its larger errors it is still (barely) consistent with the Lepage-Mackenzie
estimate.

\begin{table} \centering
\begin{tabular}{| r | l | l || l | c || l | c | }
\hline
$\beta$~ &~$W_{11}$& $\alp(\qs,W_{11})$ & $\aqs$ $r \! \geq \! a$  &
                                           $\chi^2/N_{{\rm DF}}$     &
                                           $\aqs$ $r \! \geq \! 2a$ &
                                           $\chi^2/N_{{\rm DF}}$\\ \hline
9.0  & 0.75614& 0.0731~(4)  & 0.0733~(4)   & 2.4/5 & 0.0747~(12) & 0.4/4\\
18.0 & 0.88450& 0.03040~(3) & 0.03049~(11) & 3.1/5 & 0.0311~(4)  & 0.8/4\\
\hline
\end{tabular}
\vskip 2mm
\caption{Comparison of $\aqs$ for pure SU(3) {}from our fits (fourth and
         sixth columns) with results {}from the Lepage-Mackenzie
         prescription. The latter is quoted with error $\aqs^3$.}
\label{alpLMbbig}
\end{table}

A significant change of $\aqs$ combined with a much better value for
$\chi^2$ when leaving out the first data point --- this was exactly
the kind of
behavior that raised some doubts about our estimate of $\aqs$ for
the SU(3) theory at $\beta \seq 6.8$.
But table~\ref{alpLMbbig} suggests
that our estimate of $\aqs$ {}from fits with all data points is
reliable.\footnote{Note that at fixed $r/a$ fractional lattice
artifacts do {\it not}
vanish as the lattice gets finer. So there is no obvious reason why
our method should work well on fine lattices --- meaning not just that
we get a good fit of the potential data, but also, that this involves
no significant compensating error between our handling of the lattice
artifacts and the $\aq$ extracted {}from the fit --- but not very well
on coarser ones (it should work {\it somewhat} better on fine
lattices, since the coupling is smaller).}
 Before we can draw this
conclusion with certainty, it will, however, be necessary to apply
our method to more, in particular more accurate data.
In the next subsection we perform one more
important check of our method; it allows us to
separately test our ansatz
for the lattice potential, independent of any assumption about
the form of the running coupling beyond perturbation theory.

\omitThis{
As a simple quantitative measure of lattice artifacts
let us introduce
\be\label{Rlattart}
 {\cal R} ~\equiv~ \frac{ \Va((2a,0,0)) - \Va((a,0,0))}{V(2a) - V(a)} ~-~ 1~.
\ee

\no
For pure SU(3) we find that ${\cal R}$ decreases {}from ~0.094 at
$\beta\seq 6.0$  to ~0.091 at $\beta\seq 6.8$,
{}~0.088 at $\beta \seq 9.0$ and ~0.086 at $\beta \seq
18.0$.  Clearly, it is converging to the value 0.8511 that it has for
a Coulomb potential.\footnote{Remember though, that this does {\it not}
mean
that the potential becomes more Coulomb like; the opposite is true
because of asymptotic freedom. What is true, is that as the lattice
spacing gets smaller it becomes more and more accurate to approximate
the potential {\it locally}, over a distance of $\O(a)$, by a Coulomb
potential of suitably chosen effective charge (which cancels in the
above ratio). Over a fixed {\it physical} distance the approximation
by a Coulomb potential gets worse.}
}

\vskip 1mm

To conclude this subsection, we mention that we investigated
another aspect of
the Lepage-Mackenzie prescription, namely to what extent $\aqs$ agrees
with $(2\pi)^{-4} \int_{-\pi/a}^{\pi/a} d^4 q ~\aq$. Using
$\aqs$ and $\aq$ {}from our fit, we find that the integral
is always slightly larger, but the difference is well within
the expected $\aqs^3$ error (the difference is never more than half this
error).

\subsection{The 1-Loop Expansion of the Potential}

We would now like to compare the 1-loop expansion
of our ansatz for the lattice potential with the 1-loop
% After PRDrev:
% calculation of Heller and Karsch~\cite{HelKar}
results
for the on-axis lattice potential
that can be obtained {}from
% their
the
more general results for planar
Wilson loops in finite volume
presented in~\cite{HelKar}.

Expanding the coupling of a generic scheme around $q\seq\mu$ to one loop
gives $\alp(q) = \alp(\mu) - \alp(\mu)^2 \beta_0 \ln q^2/\mu^2$.
Inserting this into eq.~(\ref{Vaofalp})\footnote{Where one should
ignore the constant $V_0$ in the following. In the end we will
consider potential differences anyhow.} we obtain
\be\label{myVaexpn}
 \Var ~=~ \alp(\mu) \Va[1](\r) ~+~ \alp(\mu)^2 \beta_0
           \Va[\ln\mu^2/q^2](\r) ~+~ \O(\alp^3) ~,
\ee
where we use the notation $\Va[\alp]$ to denote the ``potential'' in
eq.~(\ref{Vaofalp})
as a functional of the ``running coupling'' $\alp(q)$.
For any given $\alp(q)$
these functions can easily be evaluated numerically as described in
sect.~3.1. Remember  that $\Va[1](\r)$ is the lattice Coulomb
potential. The expansion in~(\ref{myVaexpn}) can easily be extended to
higher orders.

On the other hand, the 1-loop expansion of the on-axis potential given
in~\cite{HelKar} in terms of the bare coupling $\alp_0$ can be
reexpressed in terms of a continuum coupling. For the latter we choose
the V scheme.  Restricting ourselves to pure SU($N$) {}from now on, we
have (cf.~the references given above eq.~(\ref{lnWexp}))
\be\label{alpbareofalpV}
 \alp_0 ~=~ \alp_V(\mu) ~\Bigg[ 1 ~-~ \alp_V(\mu)
\Big(\beta_0 \ln\Big(\frac{1}{a\mu}\Big)^2 ~+~ 2.409830~N - \frac{\pi}{2N}\Big)
 \Bigg] ~+~ \O(\alp^3)~,
\ee
which allows us to rewrite the on-axis potential
%% of Heller and Karsch
of~\cite{HelKar}
(in the infinite volume limit) as
\bea\label{HKVaexpn}  %% 106.69806
 \Va^{(1)}(\r) = \alp_V(\mu) \Va[1](\r) +
  \alp_V(\mu)^2
 \Bigg[\beta_0 \ln\Big(\frac{a\mu}{10.3295}\Big)^2 \Va[1](\r)
 + (N^2\! - \! 1) \tilde{X}(r) \Bigg]  \nonumber\\
  +~\O(\alp^3)~.~~
\eea
Here, in terms of the notation of Heller and Karsch~\cite{HelKar},
\be\label{Xtilde}
 \tilde{X}(R) \equiv (4\pi)^2~\lim_{T\rightarrow \infty} X(R,T) - X(R,T-1)
      +\frac{1}{6}~
  \Bigg[ \overline{W}_2(R,T)^2 - \overline{W}_2(R,T-1)^2 \Bigg] ~,
\ee
which is independent of $N$.

Let us now consider the difference
$\Delta\Va \equiv a \Va((2a,0,0)) - a \Va((a,0,0))$ of the lattice
potential at the first two on-axis points.
For this quantity the optimal choice of $\mu$ according to the
Lepage-Mackenzie prescription is $\mu \seq 1.028/a$, which we will
denote by $\qsd$.
Using this value, our ansatz gives
\be\label{myVadiff}
 \Delta\Va ~=~
 C_F ~\alp_V(\qsd)~\Big[ 0.54255 ~-~ 0.08186~\beta_0~\alp_V(\qsd)\Big]
   ~+~ \O(\alp^3) ~.
\ee
%
% After PRDrev:
{}From ref.~\cite{HelKar} and unpublished results of
Heller~\cite{Helpriv} on $32^4$ lattices we estimate
%%%% $\tilde{X}(2a) - \tilde{X}(a) = 0.002097(3)$, so that
%%%% $\tilde{X}(2a) - \tilde{X}(a) = 0.3311(5)$, so that
%%%% above gives 0.234(3) below
%%                               0.002097(6) * (4Pi)^2
 $\tilde{X}(2a) - \tilde{X}(a) = 0.331(1)$, so that
\be\label{HKVadiff}
 \Delta\Va^{(1)} ~=~
 C_F ~\alp_V(\qsd)~\Big[ 0.54255 ~-~ 0.234(7)~\beta_0~\alp_V(\qsd)\Big]
 ~+~ \O(\alp^3) ~.
\ee

\begin{table} \centering
\begin{tabular}{ | c | r | l || l | l | l | l | }
\hline
Group & $\beta$~~ & $\alp_V(\qsd)$ &
           \multicolumn{4}{c|}{$\Delta \Va$}\\ \hline %% \cline{4-7}
         &               &       &
           MC & tree & our 1-loop &  1-loop\\ \hline
%6.0  & 0.262   & 0.1870(8)   & 0.190   & 0.181   & 0.175  & 0.168\\
%6.4  & 0.195   & 0.1376(8)   & 0.141   & 0.136   & 0.134  & 0.129\\
%6.8  & 0.161   & 0.1135(2)   & 0.116   & 0.113   & 0.112  & 0.108\\
%9.0  & 0.0878  & 0.0621(6)   & 0.0635  & 0.0625  & 0.0623 & 0.0611\\
%18.0 & 0.03258 & 0.02335(14) & 0.02357 & 0.02343 & 0.02342& 0.02324\\
%%
SU(3) &6.0  & 0.243   & 0.1870(8)   & 0.176   & 0.170   & 0.160\\
SU(3) &6.4  & 0.189   & 0.1376(8)   & 0.137   & 0.134   & 0.127\\
%SU(3)&6.6  & 0.172   & 0.1244(4)   & 0.125   & 0.122   & 0.116\\
SU(3) &6.8  & 0.158   & 0.1135(2)   & 0.114   & 0.112   & 0.108\\
SU(3) &9.0  & 0.0874  & 0.0621(6)   & 0.0632  & 0.0625  & 0.0611\\
SU(3) &18.0 & 0.03252 & 0.02335(14) & 0.02353 & 0.02343 & 0.02324\\
SU(2) &2.85 & 0.234   & 0.0941(2)   & 0.0951  & 0.0932  & 0.0895\\
\hline
\end{tabular}
\vskip 2mm
\caption{Results for the difference of the lattice potential
         in pure SU(3) and SU(2) at the first two on-axis points
         {}from Monte Carlo simulations compared with perturbative
         expansions in $\protect\alp_V(\qsd)$,
         where $a\qsd \seq 1.028$. The last column contains the
         1-loop result of~\protect\cite{HelKar}.}
%% HK denotes the result         of Heller and Karsch.}
\label{Vdiff}
\end{table}

\no
In table~\ref{Vdiff} we compare Monte Carlo results for $\Delta \Va$
with the tree level and 1-loop estimate {}from our ansatz and {}from
ref.~\cite{HelKar}. The errors of the Monte Carlo results were obtained
by quadratically adding those of the individual potential values, which
is almost certainly an overestimate because of correlations. To keep
all estimates strictly perturbative, we did not calculate
$\alp_V(\qsd)$ {}from our fitted $\aq$, but
instead used the Lepage-Mackenzie value of $\aqs$ {}from the plaquette
and 2-loop evolution.\footnote{If we use our fitted
$\aq$ to calculate the value of $\alp_V(\qsd)$ to be inserted in
eq.~(\ref{myVadiff}), the agreement between our 1-loop result and
the Monte Carlo data  becomes even better.
For example, in the SU(2) case we then obtain $\Delta\Va = 0.0941$.}

Note that the 1-loop estimate {}from our ansatz is consistently and
significantly better than the result of~\cite{HelKar}.  This is
somewhat odd; why should the 1-loop result {}from our ansatz be
better than what is presumably the exact answer?
Probably we are just lucky with our ansatz.
The difference between the Monte Carlo results and those
%% of Heller and Karsch
of ref.~\cite{HelKar}
is consistent with a 2-loop contribution of
{}~$1.5(2)~C_F~\beta_0^2~\alp(\qsd)^3$~ to $\Delta \Va$. The coefficient
$1.5(2)$ seems a bit large, perhaps, but it is a nontrivial fact that
this value is consistent with all data in table~\ref{Vdiff}.
We have checked that the contribution to $\Delta\Va$ {}from
zero-modes~\cite{Coste}, which was not included in~\cite{HelKar},
is negligible in the cases considered.

The situation illustrated in table~\ref{Vdiff} requires further
investigation. But we can certainly conclude that our ansatz
incorporates the lattice artifacts very well. We emphasize that
our results in table~\ref{Vdiff} do not involve any assumption about
the form of $\aq$ beyond one loop. It is therefore a check of our
ansatz for the lattice potential that is completely independent of our
ansatz for the continuum running coupling $\aq$.

\subsection{Comparing $\aq$ with 2-Loop Approximations}

We would now like to address the question to what extent the
running coupling obeys 2-loop evolution. The first question, though,
is, which form of 2-loop evolution we should use. There is $\beta_0
\alp(q) = 1/(t+b\ln(t+b\ln t))$, which incorporates the leading and
subleading 2-loop contribution. More commonly used are however the
expressions $\beta_0 \alp(q) = 1/(t+b\ln t)$, or
$\beta_0 \alp(q) = (1-b\ln t/t)/t$, which incorporate only the leading
part of the 2-loop contribution to the running coupling.  We will
refer to these expressions as the ``full'', the ``leading'', and the
``other leading'' form of 2-loop evolution of the coupling,
respectively, and generically denote them as
$\alp^{{\scsc (2)}}_{\Lam}(q)$.

We would like to compare these expressions with our fitted $\aq$,
which, though not exact, ought to be much closer to the exact answer
than any of the above expressions.  In table~\ref{alptwolpcomp} we
compare pure SU(3) fits for $\aq$ with the various forms of 2-loop
evolution. The $\Lam$-parameter used in the 2-loop expressions is
$\Lam_V$ {}from our fit.  We used the relative scales of these theories,
as determined in sect.~5 below, to give a comparison at the same
physical momenta for the different theories. The momenta are quoted in
units of the lattice spacing of the $\beta\seq 6.0$ theory.  Using
either $r_0 \simeq 0.5$~fm (cf.~sect~5)  or $\sqrt{\sig} \simeq
440$~MeV, one sees that the momenta chosen are roughly $1$~GeV,
$2$~GeV, $10$~GeV and $80$~GeV.

\begin{table}[tb] \centering
\begin{tabular}{ | l | r | l | c | c | c |}
\hline
{}~$\beta$~ & $a_{\scsc 6.0} \ts q$ & $\aq$ &
 \multicolumn{3}{c|}{$\alp_V(q)/\alp^{{\scsc (2)}}_{\Lam_V}(q)$}\\ \hline
          &                       &       &
              ~full~ & leading & ~other~\\
\hline
6.0  & 0.5   & 0.497 & 1.506 & 1.408 & 1.556\\
     & 1.0   & 0.260 & 1.186 & 1.136 & 1.244\\
     & 5.0   & 0.131 & 1.023 & 1.002 & 1.060\\
     & 40.0  & 0.0855& 1.004 & 0.993 & 1.027\\
\hline
6.4  & 0.5  & 0.497 & 1.245 & 1.150 & 1.260\\
     & 1.0  & 0.266 & 1.094 & 1.042 & 1.148\\
     & 5.0  & 0.137 & 1.011 & 0.990 & 1.050\\
     & 40.0 & 0.0884& 1.002 & 0.991 & 1.016\\
\hline
6.8  & 0.5  & 0.486 & 1.331 & 1.236 & 1.361\\
     & 1.0  & 0.258 & 1.110 & 1.060 & 1.165\\
     & 5.0  & 0.134 & 1.013 & 0.992 & 1.051\\
     & 40.0 & 0.0870& 1.003 & 0.992 & 1.026\\
\hline
$6.8'$& 0.5  & 0.493 & 1.381 & 1.285 & 1.416\\
      & 1.0  & 0.261 & 1.137 & 1.087 & 1.194\\
      & 5.0  & 0.134 & 1.016 & 0.995 & 1.054\\
      & 40.0 & 0.0867& 1.003 & 0.992 & 1.027\\
\hline
\end{tabular}
\vskip 2mm
\caption{Comparison of $\aq$ {}from our standard fit for pure SU(3)
         (using the average values {}from table~\protect\ref{Pars}) with the
         various forms of 2-loop evolution. $\Lam_V$ {}from our fit is
         used in the 2-loop formulas.}
\label{alptwolpcomp}
\vskip 1mm
\end{table}

Somewhat surprisingly, the leading form of 2-loop evolution
is closer to our fitted $\aq$ than the full one, except for the
highest momenta, even though our $\aq$ was designed to contain all
2-loop contributions. The reason for this is that up to moderate UV
momenta (around 10~GeV) there is a partial cancellation between the
subleading 2-loop and non-perturbative contributions.
Note that the other form of leading 2-loop evolution always fares
worse in our comparison than either the leading or the full
expression. This is not surprising, since compared to the full 2-loop
formula it has a subleading 2-loop term of the wrong sign. This form
of 2-loop evolution should therefore never be used, and we will ignore
it {}from now on.

While at 2~GeV the difference between our $\aq$ and 2-loop
evolution can be $10-20$\%, at 10~GeV it has already dropped to
$1-2$\%. At 80~GeV the difference is only a few permille if one
uses the the full 2-loop expression.

Finally, in figure~4 we present a graphical comparison with
2-loop evolution. We also show the values of $\alp_V(q)$
obtained at various values of $q$ using the Lepage-Mackenzie method
for various small Wilson loops and Creutz ratios. The rightmost circle
corresponds to the value extracted {}from $W_{11}$, which agrees best
with our fit, as expected.

If one wants to call the agreement between the 2-loop expressions
and our non-perturbative running coupling good or bad depends on
one's expectations. The main point is that we have a non-perturbative,
numerically accurate running coupling, and can quantify how good
2-loop evolution is.

\begin{figure}
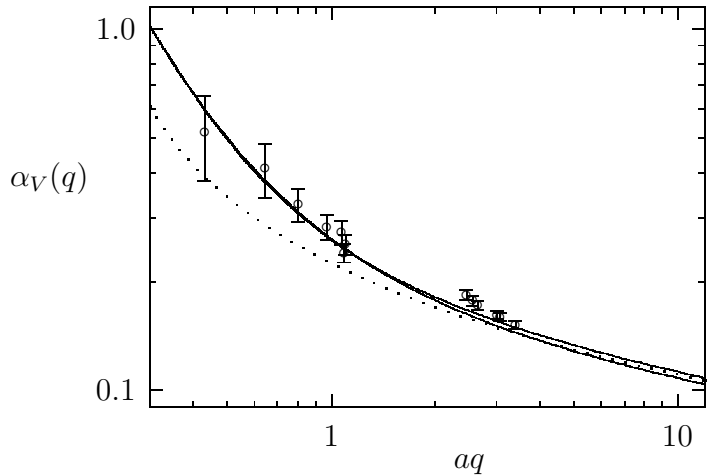
 \centering
%%\include{../../../c++/fgrid/gnuplot/alpb6.Wchi.pap}
% GNUPLOT: LaTeX picture
\setlength{\unitlength}{0.240900pt}
\ifx\plotpoint\undefined\newsavebox{\plotpoint}\fi
\sbox{\plotpoint}{\rule[-0.175pt]{0.350pt}{0.350pt}}%
%%
\caption{Log-log plot of $\alp_V(q)$ for $\beta\seq 6.0$ pure SU(3)
   {}from our standard fit (the solid lines delineating the error),
   its leading 2-loop approximation (dotted line), and results
   obtained by applying the Lepage-Mackenzie method to various
   Wilson loops and Creutz ratios (circles), shown with error
   $\alp_V(q)^3$.}
\end{figure}

\section{Scaling}

\subsection{Introduction}

By the term scaling we circumscribe three related but distinct issues
that one faces in simulating a quantum field theory on a
lattice. First, choosing and then calculating a dimensionful quantity
{}from (results of) the simulation. This quantity will then be known in
lattice units, and one can use it to determine the {\it relative}
scales of different lattice theories. Secondly, one must determine the
{\it absolute} scale of a lattice theory by equating a dimensionful
quantity {}from the simulation with its experimentally measured
value. Thirdly, one must extrapolate to the {\it continuum limit} by
simulating the theory on several (sufficiently small) lattice
spacings; or at least check that one is in the {\it scaling region}
where dimensionless ratios of physical quantities are independent of
the lattice spacing within some acceptable error.

The absolute scale determination is of course not really meaningful
unless one is sufficiently close to the continuum limit --- and to
the real world, which might involve further extrapolations,
\eg~in various masses or the number of flavors.

Note that there is no reason for the relative and the absolute scales
to be determined {}from the same quantity. In fact, this is probably not
a good idea, since it would be rather surprising if the quantity that
can be determined most precisely {}from the simulation happens to also
be the observable most suitable for precise experimental determination.
One should therefore clearly separate the absolute {}from the relative
scale determination. Once the latter has been calculated for various
theories, one just has to determine the absolute scale for
one of these theories --- using perhaps a quite different observable
--- to know it for all theories.

In the past it was customary to use the string tension $\sig$ to set
the scale. However, since it is an asymptotic (large $r$) quantity, it
has relatively large statistical and systematic errors.  Furthermore,
$\sig$ is not really well-defined in the presence of sea quarks, due
to string breaking.\footnote{Though in practice the string tension is
at present ``quite well-defined'', since clear signs of string
breaking have not been seen yet; instead one observes a large linear
regime in the static potential (cf.~the fermionic theories of
sect.~4). That such a regime should exist before the potential
flattens and keels over is also clear {}from the success of
Coulomb $+$ linear potentials in describing heavy quarkonium systems.}
Is is therefore also experimentally not particularly well determined.
The situation is even worse for the
$\Lam$-parameter, whose error is much larger, due to, as we
discovered, its correlation with higher order and non-perturbative
contributions to the running coupling at small and intermediate
distances. We now describe a better way of setting the scale.

\subsection{The Scales $q_0$ and $r_0$}

Instead of using a parameter describing the asymptotic long or short
distance behavior of the running coupling to set the scale, it is
clearly a much better idea to do so using the value of the coupling
itself at some intermediate distance.  Sommer~\cite{Som} proposed
to determine a scale $r_0 = r_0(c_r)$ using the interquark force,
\be\label{cr}
 C_F ~\alp_F(r_0) ~=~ c_r ~,
\ee
for some fixed constant $c_r$ of $\O(1)$. [For details about the coupling
in the force scheme we refer to sect.~6.1.] This is basically equivalent,
as we will see below, to defining a momentum scale $q_0 = q_0(c_q)$ via
\be\label{cq}
    C_F ~\alp_V(q_0) ~=~ c_q ~,
\ee
for suitably chosen $c_q$.

In~\cite{Som} it was suggested to take $c_r = 1.65$, since in
various phenomenological potential models of heavy quarkonium systems
this gives $r_0 \simeq 0.5$~fm, roughly the distance at which these
effective potentials are best determined empirically. Since the
relation between the static potential and these effective potentials
is {\it not} well understood at present,
%% (and the difference could be quite large~\cite{NRQCDwfn}),
$r_0$ does not seem to provide
the best way to determine the absolute scale.
The various extrapolations required to connect Monte Carlo
data to the real world
are much better understood --- in particular have been explored by
simulations --- for spin-averaged level splittings in heavy
quarkonium systems~\cite{GPLidea,SloanRev}, which are of
course also accurately known experimentally.
When available, this approach seems to
provide the best method for an absolute scale
determination (for a review of the controversy over the difference
in the absolute scale determinations {}from this method versus those
{}from light quark spectroscopy, see~\cite{SloanRev}).
Be that as it may, we will here concentrate on
determining relative scales, for which a quantity like $r_0$ is
ideally suited.

The scale $r_0 = r_0(1.65)$ has already been used in the literature,
so for comparison we will calculate it too, using the exact relation
between $\aq$ and $\aF$. {}From our point of view, it is however to
some extent more natural to define a scale $q_0$ directly in terms
of $\aq$
using eq.~(\ref{cq}). The question is if one can find a simple
relation between $q_0$ and $r_0$. In practice this is indeed the
case.  We will see in sect.~6.1 that to two loops
$\aF = \aq$ with $r={\rm e}^{1-\gamma}/q=1.526205/q$.\footnote{It is
in {\it this} sense that a momentum scale $q$ corresponds to a
distance scale $r \approx 1.53/q$, as mentioned in sect.~4. Note,
however, that one could use other prescriptions to define numerically
different momentum versus distance scale relations. In the present
context the above seems the most natural one.}
The scales we are interested in are however not in the perturbative
region, so this equality does not hold very well. Nevertheless, one
might suspect that by either
\begin{enumerate}
\item[(i)]
setting $c_q=c_r$ there is an almost universal
relation $r_0=c_{rq}/q_0$ with some $c_{rq} \neq {\rm e}^{1-\gamma}$,
or
\item[(ii)]
still requiring $r_0={\rm e}^{1-\gamma}/q_0$ there is an
almost universal constant $c_q(c_r) \neq c_r$ that achieves this.
\end{enumerate}

\no
Studying the different theories of sect.~4 we find that both
of the constants $c_{rq}$ in (i) and $c_q$ in (ii) are universal on
the permille level for $c_r = 1.65$.
We decided to basically follow approach (ii) and  define
$q_0 \equiv q_0(1.45)$. With $r_0 \equiv r_0(1.65)$ {}from now on,
we find that $q_0~r_0 = 1.524(4)$ holds for all theories
and all parameterizations of $\aq$ discussed in sect.~4.
[Considering only the standard parameterization of $\aq$ we find
$q_0~r_0 = 1.5266(12)$.]  This simple relation between $q_0$ and
$r_0$ is useful for quick translations between scale determinations
in the two schemes. Below we will however always calculate $r_0$ and
$q_0$ {}from their original definitions, given by eqs.~(\ref{cr})
and~(\ref{cq}).

%\footnote{More precisely, we find that with
%these definitions $q_0~r_0$ equals $1.5262(2)$ for pure SU(3),
%$1.5256(2)$ for pure SU(2), and $1.5275(5)$ for SU(3) with two fermions.}

\subsection{Results}

Using our standard fit and the exact relation between $\aq$ and $\aF$
we obtain the results in table~\ref{qr} for $q_0$, $r_0$
and the dimensionless quantities $q_0/\sqrt{\sig}$
and $r_0 \sqrt{\sig}$.
These estimates were obtained in the same analysis that lead to the
results in table~\ref{Pars}, including the error estimates from the
method described in  sect.~3.3. In particular,
the error of a ``composite'' quantity like $\rs$ was {\it not}
calculated  by naive error propagation from those of $r_0$ and $\sigma$,
which would ignore the correlation of these quantities.

\begin{table} \centering
\begin{tabular}{ | c | l | l | l | l | l | l | }
\hline
Group & $n_f$ & ~$\beta$~ & ~~$a q_0$ & ~$q_0/\sqrt{\sig}$ &
                           ~~$r_0/a$ & ~$r_0 \sqrt{\sig}$\\ \hline
SU(3) & 0  & 6.0  & 0.2884(8) & 1.308(6)    & 5.292(14) & 1.167(5)\\
SU(3) & 0  & 6.4  & 0.1600(6) & 1.301(10)   & 9.535(36) & 1.173(8)\\
SU(3) & 0  & 6.8  & 0.10640(13) & 1.281(4)    & 14.34(2)  & 1.190(3)\\
SU(3) & 0  & $6.8'$ & 0.1054(3) & 1.300(4)    & 14.48(5)  & 1.175(3)\\
SU(2) & 0 & 2.85  & 0.0824(8) & 1.3062(5)   & 18.54(17) & 1.1680(5)\\
SU(2) & 0 & $2.85'$ & 0.0813(11) & 1.3067(8)  & 18.76(27) & 1.1675(7)\\
SU(3) & 2W & 5.3  & 0.4225(9)  & 1.3343(18) & 3.616(8)  & 1.1448(15)\\
SU(3) & 2S & 5.6  & 0.2944(7)  & 1.3356(11) & 5.190(12) & 1.1437(10)\\
\hline
\end{tabular}
\vskip 2mm
\caption{Average values and errors of the scales $q_0$ and $r_0$
         in units of the lattice spacing and the string tension,
         obtained  {}from our standard ansatz and error analysis.
         The $\beta\seq 6.8'$ and $2.85'$ data sets are explained
         in sect.~4.1.}
\label{qr}
\vskip 3mm
\end{table}

\begin{table}[tb] \centering
\begin{tabular}{ | c | l | l | l | l | l | l | l | }
\hline
Group & $n_f$ & ~$\beta$~ & $a_1$ & ~~$a q_0$ & ~$q_0/\sqrt{\sig}$ &
                           ~~$r_0/a$ & ~$r_0 \sqrt{\sig}$\\ \hline
SU(3) & 0  & 6.0    & 0.5 & 0.2874(7)  & 1.295(9)   & 5.295(13) & 1.175(6)\\
SU(3) & 0  & 6.4    & 0.5 & 0.1589(5)  & 1.278(14)  & 9.546(33) & 1.186(10)\\
SU(3) & 0  & 6.8    & 0.5 & 0.10482(15)& 1.237(7)   & 14.37(2)  & 1.217(5)\\
SU(3) & 0  & $6.8'$ & 0.5 & 0.10471(16)& 1.262(8)   & 14.45(4)  & 1.199(6)\\
SU(2) & 0 & 2.85    & 1.5& 0.0812(9)   & 1.3076(7)  & 18.80(19) & 1.1667(7)\\
SU(2) & 0 & $2.85'$ & 1.5& 0.0802(11)  & 1.3087(10) & 19.02(26) & 1.1657(9)\\
SU(3) & 2W & 5.3    & 1.5& 0.4227(10)  & 1.3414(14) & 3.615(8)  & 1.1392(11)\\
SU(3) & 2S & 5.6    & 1.5 & 0.2938(7)  & 1.3423(12) & 5.199(13) & 1.1381(9)\\
\hline
\end{tabular}
\vskip 2mm
%\caption{Average values and errors of the scales $q_0$, $r_0$ and
%         the dimensionless combinations $\qos$, $\rs$ obtained {}from
%         $a_1 \neq 1$ fits.}
\caption{Same as table~\protect\ref{qr}, for $a_1 \neq 1$ fits.}
\label{qra}
\vskip 1mm
\end{table}

In table~\ref{qra} we show the corresponding results for our
$a_1 \neq 1$ fits (cf.~table~\ref{Parsa}).
Using the difference to our standard fits as
an indicator of systematic errors, we see that these are very small
for $q_0$ and $r_0$.  We also find that these quantities are stable
within errors when omitting
small $r$ points from the fit. The small amount they
do move, tends to brings values
from different fits closer together.
\omitThis{
This is true in particular for $\beta\seq 6.8$, where there is
a slight discrepancy between different
fits that include all small $r$ data points, cf.~tables~\ref{qr}
and~\ref{qra}.
Note that this discrepancy
is smaller for the much better $a_1\seq 0.5$
fits (any remaining small difference is perhaps due to the
slight uncertainties of the  $\beta\seq 6.8$ data themselves; remember
that in the $6.8'$ data set we omitted all points with any component
larger than $16 a$).
}
Note that the slight discrepancy in $q_0$ and $r_0$ from the $6.8$
versus the $6.8'$ data sets for $a_1 \seq 1$ disappears for the much
better $a_1 \seq 0.5$ fits.

All this is very good news: Despite the fact
that different parameterizations incorporating the 2-loop running of
the coupling lead to slight systematic errors at the ``edges'' of
$\aq$, that is, in $\aqs$ and $\sig$, they introduce no sizable
systematic bias in the intermediate region covered by the potential
data.

The dimensionless quantities $\qos$ and $\rs$, on the other hand, can
have significant systematic errors, inherited {}from $\sig$. For
the SU(2) case these systematic errors are still almost negligible,
whereas for SU(3) at $\beta\seq 6.8$ and the unquenched theories
the systematic errors are much larger than the
statistical ones.\footnote{Note, by the way,
that the statistical error of $\rs$ for the last
three entries of tables~\ref{qr} and~\ref{qra} is much smaller
than for the pure SU(3) cases, even though the opposite is true
for the errors of the potential data themselves.
We had observed and discussed similar features already in sect.~4.1}

We combine our results, including those {}from fits where small $r$
data points have been left out, and in table~\ref{rrscomp} give
estimates of $r_0$ and $\rs$ that include systematic errors. We
compare our results with estimates from various other methods: The
$\beta\seq 6.0$ and 6.4 values of $r_0$~\cite{Lufour} were obtained by
the method of ref.~\cite{Som} using force data from~\cite{BS,GBpriv};
the value of $\sig$ used in $\rs$ is from the same reference.
The $\beta\seq 6.8$ results were obtained with a modified Coulomb $+$
linear
fit~\cite{GBpriv}, as were the $n_f \seq 2$ results~\cite{HelNftwo}.
The $\beta\seq 2.85$ SU(2) value of $r_0$ was obtained in a simulation
using the Schr\"odinger functional coupling $\alp_{{\rm SF}}(q)$
{}~\cite{Divetal} (cf.~sect.~6.2).

We recall (sect.~4.1.2) that our errors for a given fit of the
$\beta\seq 2.85$ SU(2) data might not be accurate due
to correlations. However, we saw that our uncorrelated fits tend
to overestimate errors in such a situation, so all errors
in table~\ref{rrscomp} should be on the safe side.

As another check we have determined $r_0$ and $\rs$ using uncorrelated
fits to modified Coulomb $+$ linear ans\"atze. Despite the
uncertainties of the short-distance parameters of these fits, in the
intermediate range these fits are rather well-determined and we obtain
results in good agreement with the estimates from our ansatz.
% There is a somewhat subjective element in the choice of fit range,
% though.

\vskip 1mm

  We hope that an
analysis of new data at other $\beta$-values for pure SU(3) will
lead to smaller errors for $\rs$~\cite{BKS}.  Even as it stands, though,
our errors for $r_0$ and $\rs$ are much smaller than those obtained
previously (except for $\rs$ in the $\beta\seq 6.8$ case, where the
Coulomb $+$ linear estimate in table~\ref{rrscomp}
does not include all systematic errors).
%% as we know from our own Coulomb $+$ linear fits).
Note in particular that for the first time we can see a
significant difference in $\rs$ for theories with and without
dynamical fermions.

As table~\ref{rrscomp} indicates, for pure SU(3) $\rs$ scales
within errors of about 2\% for $\beta\geq 6.0$
(perhaps even for smaller $\beta$ which we have not studied so
far). Note that if we just consider our standard fit results in
table~\ref{qr},
$\rs$ even scales at the 0.5\% level.
Scaling of the string tension is shown in another way
in table~\ref{sigalpscaling}, where
we used our estimates of $\sig$  {}from table~\ref{sigma}
and rescaled them to the $\beta\seq 6.0$ lattice spacing with the
help of  the
relative scales obtained {}from  $r_0$ in table~\ref{rrscomp}.

\begin{table} \centering
\begin{tabular}{ | c | l | l || l | l || l | l | }
\hline
Group & $n_f$ & ~$\beta$~ & \multicolumn{2}{c||}{ $r_0/a$ ~{}from} &
                            \multicolumn{2}{c|}{ $\rs$ ~{}from}\\ \hline
      &       &           &  ~our fits &  ~other  &
                             ~our fits &  ~other\\ \hline
SU(3) & 0  & 6.0  & 5.296(16) & 5.44(26)   & 1.172(11)     & 1.23(7)\\
SU(3) & 0  & 6.4  & 9.54(4)   & 9.90(54)   & 1.180(16)     & 1.16(6)\\
SU(3) & 0  & 6.8  & 14.42(7)  & 14.36(8)   & 1.197(25)     & 1.206(8)\\
SU(2) & 0 & 2.85  & 19.0(4)   & 20.6(14)   & 1.167(2)      &\\
SU(3) & 2W & 5.3  & 3.624(14) & 3.7(2)     & 1.140(5)      & 1.16(6)\\
SU(3) & 2S & 5.6  & 5.201(15) & 5.2(2)     & 1.140(3)      & 1.14(4)\\
\hline
\end{tabular}
\vskip 2mm
\caption{Comparison of determinations of $r_0$ and $\rs$ from our
         and other methods.
         The references for the latter are given in the main text.}
\label{rrscomp}
\vskip 5mm
\end{table}

\begin{table} \centering
\begin{tabular}{ | l | l | l | }
\hline
{}~$\beta$ &   ~~$a^2_{{\scsc 6.0}} ~\sig$ &
             ~~$\alp_V(q^\star_{{\scsc 6.8}})$\\ \hline
6.0      &   ~0.0491(8)  & ~0.1114(23)\\
6.4      &   ~0.0496(13) & ~0.1164(17)\\
6.8      &   ~0.0519(22) & ~0.1142(8)\\
\hline
\end{tabular}
\vskip 2mm
\caption{String tension for pure SU(3) in units of the $\beta\seq 6.0$
         lattice spacing, and $\aq$ evolved to the common scale
         $q^\star_{{\scsc 6.8}}$.}
\label{sigalpscaling}
\vskip 1mm
\end{table}

In this table we also show, as check of UV scaling, results for $\aq$
at the scale $q^\star_{{\scsc 6.8}}\seq 3.4018/a_{{\scsc 6.8}}$ for
different $\beta$. [A rough check of UV scaling, without error bars,
was already provided by the $\aq$ entries in table~\ref{alptwolpcomp}.]
We used our $a_1 \seq 0.5$ fits and evolved the $\beta\seq 6.0$ and 6.4
results using the appropriate ratios of $r_0$. The error {}from the
scale uncertainty is negligible. Instead, the error is dominated by the
error of our fit (where one must be careful to take into account the
correlation of the fit parameters) and by the systematic error of
$\aqs$, which we took to be the difference between the central values
in tables~\ref{Pars} and~\ref{Parsa} (for $\beta\seq 6.8$ we averaged
over the 6.8 and $6.8'$ results). We see that scaling violations, if
existent, are just slightly larger than the error of $1-2$\%.

Without presenting explicit results here, we note that scaling holds at
the same or even slightly better level of
accuracy also in the intermediate
regime\footnote{One should not be worried about the
fact that the crossover parameter $c$ does not scale too well (for given
$a_1$; it obviously should not scale when comparing results with
different $a_1$). It is an {\it effective} parameter,
representing small subleading effects, and therefore prone to amplify
even small flukes in the data.}
(to some extent this can be seen, again, in table~\ref{alptwolpcomp}).
We hope to provide more
accurate results for pure SU(3) in~\cite{BKS},
including a more stringent check of the scaling of $\aq$ at all
momenta.

Finally, we have repeatedly stressed the large systematic errors
of the $\Lam$-parameter, that are usually ignored. To underscore
this point, we compare in table~\ref{LamComp} three different
determinations of this parameter: First the value {}from our fit,
second that obtained by the Lepage-Mackenzie prescription {}from the
plaquette by assuming (leading) 2-loop evolution
(cf.~table~\ref{WalpLM}), and thirdly that obtained by matching force
data to the 2-loop $\aF$.
We used $\Lam_V \simeq 1.526 \cdot\Lam_F$ to translate {}from the force to
the V scheme (cf.~sect.~6.1).
Note that our result is the only one taking into account
the most important systematic errors,
which we estimated {}from tables~\ref{OptPars} to~\ref{Parsa}.
Our estimate is systematically lower, but agrees quite well
with that obtained {}from the plaquette. As a somewhat ``old-fashioned''
check of scaling we tabulate our estimate of $\Lam_V/\sqrt{\sig}$
in the last column. Within large errors scaling seems to hold
also for this quantity.

\begin{table} \centering
\begin{tabular}{ | c | l | l || l | l | c || c | }
\hline
Group & $n_f$ & ~$\beta$~ & ~~$a \Lam_V$ &
        $a\Lam_V^{{\scsc (2)}}(W_{11})$ &
        $a\Lam_V$ (force) & $\Lam_V/\sqrt{\sig}$\\
\hline
SU(3) & 0  & 6.0  & ~0.13(2)    & ~0.169(13) & 0.220(7) & 0.59(9)\\
SU(3) & 0  & 6.4  & ~0.091(8)   & ~0.097(6)  & 0.119(4) & 0.74(7)\\
% SU(3) & 0  & 6.8  & ~~0.056(3)   & ~0.058(3)  & 0.070(8) & 0.67(4)\\
SU(3) & 0  & 6.8  & ~0.056(3)   & ~0.058(3)  &          & 0.67(4)\\
SU(2) & 0 & 2.85  & ~0.051(3)   & ~0.055(7)  & 0.063(5) & 0.83(5)\\
\hline
\end{tabular}
\vskip 2mm
\caption{Comparison of our, the Lepage-Mackenzie, and the estimate
         of $\Lam_V$ {}from force data using the 2-loop
         $\aF$~\protect\cite{Bali,UKsutwo}.
         Our value of $\Lam_V$ includes an estimate of
         systematic errors.
         The last column is our estimate of $\Lam_V$ in units of the
         string tension.}
\label{LamComp}
\vskip 5mm
\end{table}

\section{Other Schemes for the Running Coupling}

We now briefly discuss two of the more established methods of obtaining
the running coupling in different schemes. Many other schemes have
recently also begun to be explored, see~\eg~\cite{TP,Divetal,Parri}.

\subsection{The Force Scheme}

We already mentioned this scheme at various points in this paper. We now
discuss it in more detail.  Let us first describe the close relation of
the coupling in this scheme, $\aF$, to $\aq$. Recall that $\aF$ is
defined in terms of the continuum potential as
\be\label{alpFdef}
 r^2 V'(r) ~\equiv~ C_F \alp_F(r) ~.
\ee
So just {}from their definitions we have an explicit exact relation
between these two couplings. This relation can be rewritten as
%% DO I NEED THIS ??
%
 \be\label{alpFfromV}
 \alp_F(r) ~=~ \frac{\sig}{C_F} ~r^2 ~+~ \frac{2}{\pi} \int_0^{\infty}
 ~\frac{dq}{q} ~\sin q r ~
 \big[ q \partial_q \tilde{\alp}_V(q) + \tilde{\alp}_V(q)\big ] ~,
 \ee
where the ``IR subtracted'' $\aq$ is defined by
 \be\label{subtralpV}
 \tilde{\alp}_V(q) ~=~ \aq ~-~ \frac{2 \sig}{C_F}~\frac{1}{q^2} ~.
 \ee

Besides this exact though perhaps not directly illuminating relation,
there is a ``deeper'' relationship between these two couplings.
Namely, $\aF$ is in a rather literal sense a position space analog
of $\aq$. To see this, note first of all that, like any continuum
coupling, $\aF$ obeys the $\beta$-function equation with the first
two universal coefficients. In particular, it has an expansion
of the form~(\ref{alpoftexpn}), where now
$t\equiv \ln(\Lam_F \ts r)^{-2}$.  The  relation between
$\Lam_F$ and $\Lam_V$
can be obtained by a direct calculation of the 1- and 2-loop
terms in $V(r)$, written as a Fourier transform of $\aq$. We obtain
{}~$\Lam_F={\rm e}^{\gamma-1} \Lam_V$, where $\gamma=0.57721566\ldots$
is Euler's constant.

Next, note that for long distances
 \be
 \alp_F(r) ~=~ \frac{\sig}{C_F} ~r^2 ~+~ \frac{\eIR}{C_F} ~+~\ldots ~,
 \ee
which is to be compared to the IR expansion~(\ref{alpIR}) of $\aq$.
So we see that
one can use the same ansatz~(\ref{myalp}) for $\alp_F(r(t))$ as for
$\alp_V(q(t))$.

We mentioned in the introduction that the force scheme has been widely
used~\cite{fitMich,UKsutwo,BS,Bali,UKforce}
to obtain an estimate of the $\Lam$-parameter by matching force data
to the (leading) 2-loop expression for $\aF$. In sect.~5 we also
discussed how $\aF$ is used to define the useful intermediate scale
$r_0$ by $C_F \alp_F(r_0) \seq 1.65$. Sommer~\cite{Som} determined
$r_0$ {}from force data.
Let us note some potential and real problems in using
the force scheme to determine $\Lam_F$ or $r_0$, and to what extent
some of these problems can be circumvented.

\begin{enumerate}

\item[$\bullet$]
 To get the force {}from potential data involves taking numerical
derivatives, which generically increases the errors dramatically.
Only if the potential data at neighboring points are strongly
correlated will this not be the case.

\item[$\bullet$]
 Lattice artifacts in the force can easily be taken into account
at tree level, cf.~\cite{Som}. However, we know that at short distances
this is not sufficient to describe the lattice artifacts accurately.
For the $r_0$ determination this will presumably lead to significant
errors only, if at all,
 on rather coarse lattices, where $r_0/a$ is not much larger
than~1, cf.~table~\ref{rrscomp}. But for the $\Lam_F$ estimate this
will induce sizable systematic errors even on fine lattices.

\item[$\bullet$]
 Global versus local fits: To determine $r_0$ {}from force data one
only needs a locally accurate interpolation between two neighboring
points~\cite{Som}. $r_0$ can therefore be obtained without any
assumption about the global {}from of the force. By the same token,
however, one misses the opportunity of obtaining a
more precise value for $r_0$ offered by a globally accurate ansatz (or
accurate at least in the intermediate region) that takes into
account information {}from many more data points. For the $\Lam_F$
estimates one often has the worst of both worlds, namely, $\Lam_F$ is
typically extracted locally {}from the force using the (leading) 2-loop
formula for $\aF$, which can not be expected to hold very
accurately. This problem in itself could be mitigated by fitting to a
global ansatz like~(\ref{myalp}). However, lattice artifacts would still
be a problem.
\end{enumerate}

\no
Note that  none of the above problems occurs in our approach: We
have a globally accurate\footnote{Recall that we checked that leaving
out points at small
or large distances from the fit does not significantly change our
estimate of $r_0$.}
ansatz for the potential that takes lattice artifacts into account much
better than at tree level. What we have not done yet, is to take into
account correlations of the potential data. If these were
strong, so that the force could be extracted with great accuracy, the
best method to fit the force data would presumably be to match them to
the corresponding differences of the lattice potential as given by our
ansatz. This would solve the one problem that our approach has, namely
that we have to fit the overall constant of the potential, whose
correlation with $\aqs$ provides the main contribution to the latter's
error.

However, in most cases the correlations of the potential data do not
seem to be strong enough to outweigh the other disadvantages of using
the force. And, in any case, since the potential data are primary even
in the presence of correlations, the cleanest method of extracting a
running coupling would be to append our fitting routine to that for
the lattice potential and perform fits on bootstrap copies of the
potential data. This would automatically take care of all
correlations, presumably in the best way possible.

\subsection{The Schr\"odinger Functional Coupling}

 L\"uscher {\it et~al}~\cite{Lutwo,Luthree,Lufour} have
introduced and studied in some detail a completely different
running coupling. It is defined by the response of the partition
function (``Schr\"odinger functional'') of a lattice gauge theory
on a finite lattice to a constant background field, with the
extent of the lattice providing the running scale. This coupling
is denoted by $\alp_{{\rm SF}}(q)$. The great advantages of this
method are (i) that the coupling is defined so that it can be
extracted {}from the Monte Carlo simulation with basically no
systematic error and (ii) that it is relatively easy to cover
a large energy range by varying the size of the lattice. Its
disadvantages are (i) that it involves ``special purpose'' simulations
due to the gauge field configurations required and (ii) that there
are apparently large higher order coefficients relating this
scheme to more common ones, see below (the 2-loop relation to the
$\MSbar$ scheme will be known soon~\cite{LVW}).

\subsection{A Comparison}

We would like to end this section with a quantitative comparison
of $\alp_F$, $\alp_V$ and $\alp_{{\rm SF}}$ at a suitable physical
scale.
For pure SU(3) at $\beta\seq 6.5$ the most accurate value of $\alp_F$
obtained by the UKQCD collaboration {}from force data~\cite{UKforce}
turns out to be at $r/a \seq  2.5322$, where they find
$\alp_F \seq 0.248(2)(1)$. On the other hand, translating results for
$\alp_{{\rm SF}}(q)$ via perturbative matching
formulas gives $\alp_F \seq 0.205(7)(9)$ at the same physical
scale~\cite{Lufour} (we quote $\alp^3_F \simeq 0.009$ as the
second, systematic error {}from the perturbative matching).
We do not have results at
$\beta\seq 6.5$.  But using $r_0/a \seq 11.23(21)$ for
$\beta\seq 6.5$
(from~\cite{Lufour}, obtained  using data of~\cite{UKforce})
%% ~\cite{CMquoted}
and our values of $r_0$ we obtain
$\alp_F \seq 0.231(5), ~0.237(5), ~0.231(4)$ at the same physical
scale for $\beta\seq 6.0, ~6.4, ~6.8$, respectively. In our
error estimate
we linearly added the contributions {}from our
$\aq$ and those {}from expressing the above scale $r$ in the appropriate
lattice units.
The systematic error {}from the $a_1\seq 1$ versus $a_1 \seq 0.5$ fits
is negligible.
 The non-monotonicity of the $\alp_F$-values as a
function of $\beta$, which we previously saw for $\aq$
(cf.~tables~\ref{alptwolpcomp} and~\ref{sigalpscaling}), is perhaps
due to systematic errors of the potential data we used in
our fits. We therefore simply quote $\alp_F \seq 0.236(6)$ at the
above scale. This is in reasonable agreement with the value {}from
force data, but they both differ by about $3\alp^3$ with respect
to the value {}from the Schr\"odinger functional. This indicates that
perturbative corrections in the translation between $\alp_{{\rm SF}}$
and $\alp_V$ or $\alp_F$ have quite large coefficients.

\section{Conclusions and Outlook}

Our ability to fit lattice potential data to very high precision
rests on two ingredients. First, our expression for the lattice
potential in terms of the continuum running coupling $\aq$,
eq.~(\ref{Vaofalp}), and, secondly, our parameterization of
$\aq$ that incorporates both the short distance QCD
and the long distance flux tube predictions for the static
potential (sect.~2).

Our ansatz for the lattice potential is obviously correct at tree
level.
At the 1-loop level we found (sect.~4.3),
rather confusingly, that our ansatz seems
to work much {\it better} than what is presumably to be the exact
answer~\cite{HelKar}.  This remains to be understood. Be that as it
may, this and other checks, like the comparison with the
 Lepage-Mackenzie estimate of the short-distance coupling (sect.~4.2),
indicate that we are handling the lattice
artifacts extremely well. More precisely, the fact that we can reproduce
basically all lattice artifacts within the very small errors of the
potential data does not seem to involve any significant compensating
error between our ansatz for the lattice potential and our estimate
of the running coupling $\aq$ at short distances. To be absolutely
sure about this conclusion, however, we do need new potential data,
since
at the level of precision we are working many presently available
data seem to suffer {}from slight systematic uncertainties of
one sort or another (finite-size effects, extrapolation of Wilson loop
ratios to large euclidean times).

Previous methods of fitting the
data did not incorporate QCD very well
--- to the extent that (slight) systematic errors of the data did not
make much of a difference; they could not be detected  anyhow.
%
%\footnote{A dramatic illustration of this fact is
%provided by the $\beta \seq 6.8$ pure SU(3) data of~\cite{BS}.
%A modified Coulomb $+$ linear  fit yielded~\cite{GBpriv}
%$\chi^2/N_{{\rm DF}} \seq 126/67$ after leaving out the first two
%data points. Our fit gave $\chi^2/N_{{\rm DF}} \seq 5.9/69$. Clearly
%the data are strongly correlated, but this fact could not be detected
%with the  Coulomb $+$ linear fit, apparently
%due to the systematic errors of the ansatz itself.
%[These data, which are {\it not} the $\beta\seq 6.8$ SU(3) data used
%in this paper,  turned out to also suffer {}from large finite-size
%artifacts, but that is besides the point here.]}
%
Our method is precise enough to make it necessary to carefully
check the data for finite-size effects and extrapolate the effective
potentials to large euclidean times. Put in another way, there is
a large payoff again for precise potential data.
Particularly exciting is the prospect of potential data {}from improved
actions~\cite{improv}, for which our method is
ideally suited (cf.~the introduction), since what eventually will
prevent us {}from using our method
to extract  $\aq$ ever more accurately,
are lattice artifacts, and not, for example,
a 2-loop approximation used in other methods.

We believe that performing fits of potential data with our method
allows for the first time a {\it realistic} error assessment,
in contrast with the hard to estimate systematic
errors afflicting (modified)
Coulomb $+$ linear fits and the method of estimating
the $\Lam$-parameter by matching force data to 2-loop formulas.

We showed that the $\Lam$-parameter has a relatively large error
(cf.~sect.~4.1.1) --- without leading to a large error in $\aq$ itself
--- because $\Lam$ is strongly correlated with higher order and
non-perturbative contributions at an intermediate range, appearing
here in the form of the crossover parameter $c$. This is not
surprising, if one remembers that the true UV regime is $q \gg \Lam$,
\ie~$\Lam$ is really more an intermediate range parameter; it must be
correlated with higher order effects.  We therefore regard previous
estimates of $\Lam$ as overly optimistic.
% since no real effort was made to estimate systematic errors.

Within our approach the error of $\Lam$ is largely a red herring. We
only use $\Lam$, which is clearly a bad way to parameterize the physics,
because it is otherwise difficult to write down a closed form expression
for the running coupling. But the physics is contained in
$\aq$, which has much smaller statistical and systematic errors.

The scale $r_0$ is now widely used to compare the lattice spacings of
different theories. For such relative scales, this is certainly much
more accurate than using the $\Lam$-parameter or the string
tension. Sommer~\cite{Som} described a careful procedure, taking
lattice artifacts into account at tree level, to determine $r_0$
locally {}from force data. We determine $r_0$ {}from our fit of
potential data, using the exact relation between the force and
potential schemes. Our values for $r_0$ (sect.~5) are significantly
more accurate than those {}from the procedure of ref.~\cite{Som} ---
or those obtained using a modified Coulomb $+$ linear ansatz for the
potential --- because (a) we use potential data, that in the cases
considered have much smaller errors than force data, (b) we can take
lattice artifacts into account more accurately, and (c) we have a
globally accurate ansatz for the potential, so that information {}from
many points, not just two neighboring ones, can be taken into
account.\footnote{We carefully checked that the one potential problem
with global fits, systematic errors {}from the form of the ansatz,
does not lead to significant errors in the intermediate region where
$r_0$ is situated.}

Using $r_0$ to set the (relative) scale of different lattice theories,
enabled us to check the scaling of $\aq$ for pure SU(3),
cf.~sect.~5.3.  If existent, the scaling violations are not much
larger than the error, which is about 2\%. Unfortunately, the pure
SU(3) data on the finest lattice, corresponding to $\beta\seq 6.8$,
show some signs of systematic errors, making a more accurate check of
scaling impossible, since we investigated only two other $\beta$
values, $6.0$ and $6.4$. A more complete study of SU(3) data is in
preparation~\cite{BKS}, and will provide a more stringent test of
scaling.

\vskip 1mm

When comparing quenched and unquenched QCD we clearly see that the
effect of adding fermions can {\it not} be absorbed into a
``$\beta$-shift''. If one matches the string tensions in this manner,
one finds that the running coupling in the UV is quite different
(sect.~4.1.1).
Also, our results are accurate enough to show, for the first time,
that the dimensionless quantity $\rs$ is different
in quenched and unquenched QCD (sect.~5.3).

\vskip 1mm

Till now we performed only uncorrelated fits, ignoring the
correlations between the potential at different $\r$-values.
Except for the case of the SU(2) potential~\cite{UKsutwo}, the data
we used are not strongly correlated.
In such cases
previous experience with (modified) Coulomb $+$ linear
fits~\cite{GBpriv,Helpriv,corrfits} indicates,
that neither the fit parameters nor their errors change
significantly when employing correlated fits. We have explicitly
checked this (sect.~4.1.2) by performing fits of the potential data
to modified Coulomb $+$ linear ans\"atze and comparing them with
previous results from correlated fits. For strongly correlated data
we find that our naive uncorrelated fit {\it over}- not underestimates
the errors (what uncorrelated fits {\it do} underestimate,
is $\chi^2$). It is easy to understand why this is true.
So in all cases our errors seem to be on the safe side.

However,
what clearly should be done, is to merge our fitting routine with that
for the potential and perform fits on bootstrap copies of
the potential; this would automatically take into account all
correlations in an optimal manner and simplify the error analysis.
Note that once this has been done, another nice feature of our approach
can be fully exploited. Namely, that our method just needs one fit of
potential data to obtain the running coupling $\aq$ at all (accessible)
scales; in particular, we obtain estimates of the UV coupling $\aqs$
(or $\Lam_V$, if desired), the string tension $\sig$, and $r_0$ in one
fell swoop. At present one usually performs a separate fit (or even
simulation) for each of these quantities.

\vskip 1mm

We know how to incorporate QCD to any number of loops in our
ansatz for the running coupling (sect.~2.1).
 Once the third $\beta$-function coefficient, $\beta_2$,
becomes available in the V scheme it should of course be incorporated
into our ansatz.
With present data we can not determine this parameter reliably by
fitting. It is suggestive, though, that pure SU(3) on finer
lattices seems to favor a value of $a_1$ (which is related to
$\beta_2$) that is  in the neighborhood of the value this
quantity has in the $\MSbar$ scheme (cf.~sect.~4.1). With future
high precision data it therefore might be possible to determine
$\beta_2$ by fitting. On the other hand,
non-perturbative effects might become important
before 3-loop effects are clearly visible.

\vskip 1mm

Besides fitting lattice potential data and extracting the running
coupling along the way, our results should have various other
applications in the upcoming era of high precision Monte Carlo
calculations. For example, Monte Carlo simulations are now
competing (see~\cite{SloanRev} for a recent review)
with --- or have already surpassed ---
experiment in determining the running coupling of real world QCD at
high energy (this is due to the fact that in the simulation one
can use a more suitable observable, like level splittings in
bottonium~\cite{GPLidea,SloanRev}, to set the scale  than is possible
in accelerator experiments).
To minimize the systematic error in these calculations
it is crucial to be able to extrapolate the running coupling to
different energy scales and to a number of fermions different {}from
those used in a specific simulation.

Another obvious application is to potential models (on the lattice or
in the continuum). As far as we know, previously no simple analytic
expression incorporating both 2-loop QCD and the leading and
subleading string picture prediction for the potential was
known. Although the relation between the static potential and heavy
quark potentials is {\it not} well understood, in high precision
studies of the heavy quark spectrum our form of the potential might
also present an improvement.  More generally, our parameterization of
the running coupling should be useful for just about any phenomena
where one would like to extend perturbation theory without running
into the Landau pole.

\vskip 1mm

Finally, we note that our scheme can     accommodate string breaking
in the presence of light quarks. If the onset of string breaking
(before the Yukawa-like potential between the pair-created mesons
becomes relevant) is modeled by a potential
$V(r) = \frac{\sig}{\mu} (1-{\rm e}^{-\mu r})$ for large $r$,
the small $q$ behavior of the running coupling is
\be
\alp_V(q) ~=~\frac{2}{C_F}~\frac{\sig q^2}{(q^2 +\mu^2)^2} ~+~\ldots~.
\ee
To take this into account we add a suitable term to eq.~(\ref{myalp}),
\be
 \frac{1}{\beta_0 \ts \aq} ~=~
  \frac{C_F \ts \mu^4}{2 \sig \beta_0}~\frac{1}{q^2} ~+~
   \frac{C_F \ts \mu^2}{\sig \beta_0}~\frac{1}{1+q^4/\Lam_V^4} ~+~
 \ln\Big[1~+~\frac{q^2}{\Lam_V^2}~\ln^b\Big(c_0+\frac{q^2}{\Lam_V^2}
 \lam(q)\Big) \Big] ~,
\ee
where there is some freedom in the exact choice of the second term on
the rhs.

\vskip 1mm

Another, less trivial extension would be to generalize our ansatz for
the infinite lattice potential to one for a finite lattice. It is well
known, at least empirically, that the right potential on a finite
lattice is {\it not} obtained {}from some infinite lattice expression
by simply replacing the integral over the Brillouin zone by a
corresponding sum over the lattice.
% For distances that are not very
% small with respect to the lattice size, such sums give very wrong
% answers.
In understanding this problem it will be necessary to take
into account the exact definition of the finite-volume potential,
\eg~via Wilson loops, a consideration that played no role in the
present work.

\bigskip

\section*{Acknowledgments}

I would like to thank Urs Heller and Chris Michael for access to their
data (and Andrew Lidsey for sending me some of them).  I am very
grateful to Gunnar Bali for providing me with various preliminary and
the final analyses of the $\beta\seq 6.8$ pure SU(3) data before
publication, and for helpful comments on the manuscript.  Finally, I
would like to thank the above, as well as Mark Alford, Kent
Hornbostel, Rainer Sommer, Henry Tye, and especially Peter Lepage for
many useful discussions.  This work is supported by the NSF.

\end{document}